\documentclass[twocolumn]{aastex62}

\usepackage{appendix}
\usepackage{amsmath,amssymb, amsthm,amstext}
\usepackage{natbib}
\usepackage{graphicx}
\usepackage{color}
\usepackage{array, enumerate, mathtools}
\usepackage{bm}
\usepackage{multirow}
\usepackage{braket}
\usepackage{txfonts}

\newcommand{\pder}[2][]{\frac{\partial#1}{\partial#2}}

\def\be{\begin{equation}}
\def\ee{\end{equation}}
\def\risco{r_{\rm isco}}
\def\dotJI{\dot J_{\rm mig, I}}
\def\Medd{\dot M_{\bullet}^{\rm Edd}}
\def\medd{\dot m_{\bullet}^{\rm Edd}}
\def\mdoto{\dot m_{\rm in}(r_{\rm obd})}

\submitjournal{ApJ}
\shorttitle{Supercritical accretion of stellar-mass compact objects in active galactic nuclei}
\shortauthors{Pan \& Yang}

\begin{document}
\title{Supercritical accretion of stellar-mass compact objects in active galactic nuclei}

\author{Zhen Pan}
\affiliation{Perimeter Institute for Theoretical Physics, Waterloo, Ontario N2L 2Y5, Canada}
\email{zpan@perimeterinstitute.ca}

\author{Huan Yang}
\affiliation{Perimeter Institute for Theoretical Physics, Waterloo, Ontario N2L 2Y5, Canada}
\affiliation{University of Guelph, Guelph, Ontario N1G 2W1, Canada}
\email{hyang@perimeterinstitute.ca}

\begin{abstract}
  Accretion disks of active galactic nuclei (AGN) have been proposed as promising sites for producing
  both  (stellar-mass) compact object mergers and extreme mass ratio inspirals. Along with the disk-assisted migration/evolution process, ambient gas materials  inevitably accrete onto the compact objects. The description of this process is subject to significant theoretical uncertainties in previous studies.
  It was commonly assumed that either an Eddington accretion rate or a Bondi accretion rate (or any rate in between) takes place, although these two rates can differ from each other by several orders of magnitude. As a result, the mass and spin evolution of compact objects within AGN disks are essentially unknown.  In this work, we construct a relativistic supercritical inflow-outflow model for black hole (BH) accretion. We show that the radiation efficiency of the supercritical accretion of a stellar-mass BH (sBH) is generally too low to explain the proposed electromagnetic counterpart of GW190521. Applying  this model to  sBHs embedded in AGN disks,
  we find that, although the gas inflow rates at Bondi radii of these sBHs are in general highly super-Eddington,
  a large fraction of inflowing gas eventually escapes as outflows so that only a small fraction  accretes onto the  sBH,
 resulting in mildly super-Eddington BH absorption in most cases.
  We also implement this inflow-outflow model to study the evolution of
  neutron stars (NS) and white dwarfs (WD) in AGN disks,
  taking into account corrections from star sizes and star magnetic fields.
 It turns out to be difficult for WDs to grow to the Chandrasekhar limit via accretion because WDs are spun up more efficiently to reach the shedding limit before the Chandrasekhar limit. For NSs the accretion-induced collapse is possible if NS magnetic fields are sufficiently strong to guide the accretion flow outside the NS surface,
   keeping the NS in a slow rotation state during the whole accretion process.
\end{abstract}
\keywords{Active galactic nuclei(16); Black holes(162); Bondi accretion(174); Gravitational waves (678)}

\section{Introduction}
With a growing catalog of bianary black holes, binary
neutron stars and black hole-neutron star systems detected by Advanced  LIGO and Virgo,
in recent years extensive studies have been conducted on their possible formation channels  \citep{LVC2021}.
In addition to isolated binaries \citep[e.g.][]{Zwart2000,Hurley2002,Belczynski2002} and dynamical formation in dense stellar clusters \citep[e.g.][]{Sigurdsson1993,Lee1995,Zwart2000,Banerjee2010,Stephan2016,Bartos2017},
binary formation in accretion disks of active galactic nuclei (AGN) has been proposed as
another promising and interesting channel \citep{Stone2017,Bartos2017,McKernan2018,Leigh2018,Yang2019,Yang2019prl,Yang2020,Yang2020b,Secunda2019,Secunda2020,
McKernan2020,McKernan2020b}. The gas-rich environment is expected to leave unique observational signatures on mergers of
stellar-mass compact objects in AGN disks. For example,
binary black hole (BBH) mergers in AGN disks
may generate detectable electromagnetic (EM) signals \citep{McKernanFord2019,Graham2020,Kimura2021,Wang2021}
which should be absent in other formation channels; gamma-ray bursts arising from
binary neutron star (BNS) mergers in AGN disks may also yield different EM signals and detectable high-energy
neutrino emission \citep{Kimura2021,Perna2021a,ZhuWang2021b,ZhuWang2021a,ZhuZhang2021}.
During the entire inspiral stage, gas accretion onto these compact objects may change their masses and spins,
e.g., producing compact objects within the lower and upper mass gaps, and compact binaries with non-zero effective spins  \citep{McKernan2020b, Yang2020,Tagawa2020b,Tagawa2020,Tagawa2021,Gilbaum2021}.

Besides forming binaries and eventually produce coalescence gravitational wave signals in the frequency band of ground-based detectors, compact objects  embedded in AGN disk give rise to plenty of other interesting phenonmena. For example,
stellar-mass BHs (sBHs) may be captured by AGN disks and  then migrate inward to the galactic-center, massive BH (MBH), which has been suggested as
an important or even dominant formation channel of extreme mass ratio inspirals (EMRIs) \citep{Pan2021,Pan2021b} based on the rate calculations, so that they  belong to the primary targets for
the laser interferometer space antenna (LISA) with possible environmental waveform characteristics \citep{Bonga:2019ycj,Yang:2017aht,Yang:2019iqa,Yunes:2011ws}; the birth and growth of intermediate-mass black holes
in AGN disks have been investigated \citep{McKernan2012,McKernan2014};
  observation prospects of accretion onto compact objects and accretion induced collapses of neutron stars (NSs)
 and white dwarfs (WDs) in AGN disks have also been discussed in detail \citep{Wang2021a,Perna2021,ZhuYang2021}.

On the other hand, the accretion onto compact objects in AGN disks plays an essential role in determining the mass and spin evolution of these objects. Straight estimation using the  Bondi accretion  formula (without considering possible outflows) suggests that the  accretion is highly supercritical,
the  rate of which is usually many orders of magnitude higher than the Eddington accretion rate.
In previous  studies  \citep[e.g.,][]{McKernan2020b, Yang2020,Tagawa2020b,Tagawa2020,Tagawa2021,Gilbaum2021},
an Eddington accretion rate or a Bondi accretion rate is  usually assumed,
which makes model predictions that are sensentive to the accretion rate highly uncertain.
As shown in many analytic models \citep{Blandford1999,Blandford2004,Begelman2012} and numerical simulations \citep{Yuan2012b,Yuan2012a,Sadowski2014,McKinney2014,Yang2014,Takahashi2018}, outflow naturally
emerges in cases of supercritical accretion of compact objects, such that only a (small) fraction of inflowing gas at
the outer boundary finally accretes onto the central compact object.
In this paper, we aim to build a relativistic supercritical inflow-outflow model of BHs and apply this model
to study the mass and spin evolution of sBHs embedded in AGN disks, along with EM emissions.
Given a mass inflow rate at outer boundary, all the flow properties, including the disk structure, radiation cooling rate
and the accretion rate onto the  sBH, are uniquely determined within the supercritical inflow-outflow model.
We find supercritical accretion flows are in general radiation inefficient and advection dominated, similar to the well-known
advection dominated accretion flows (ADAFs) \citep{Narayan1994,Narayan1997}. In particular, it is unlikely to explain
the proposed bright EM counterpart of GW190521, if it indeed orignated from the same BBH system
\citep[see also][]{Ashton2020,Nitz2021,Palmese2021},  by supercritical accretion of the remnant BH
in an AGN disk as claimed in \cite{Graham2020}.

After applying this supercritical accretion model on sBHs embedded in AGN disks,
we find that the gas inflow rate at the outer boundary $r_{\rm obd}$, where the infalling gas circularizes into a disk profile,
is in general highly super-Eddington, i.e., $\dot m_{\bullet}(r_{\rm obd})\gg \medd$. A large fraction of inflowing gas
escapes as outflow and only a small fraction gets accreted by the central sBH, and the accretion onto sBHs turns out to be mildly super-Eddington in most cases, i.e., $\dot m_{\bullet, 0}\gtrsim \medd$. As a result,  the majority of sBHs that are captured onto the AGN disk and migrate into the MBH within the AGN disk lifetime only grow by a small fraction due to accretion, and the majority of the mass accumulation happens in the vicinity of the central MBH,
where the gas density is high.
We also calculate the spatial distribution of sBHs captured into the disk, which suggests that $\approx (5-50)$ sBHs aggregate around the MBH within $10^3$ times the MBH gravitational radius, depending on the disk model and the disk lifetime. It has an interesting implication
for sBHs in Sgr A*: a number of sBHs should be brought to the vicinity of the galactic MBH during its previous AGN phase,
$\approx (2-12)$ of which are expected to be still orbiting around the Sgr A* today if the AGN phase happened in less than $10$ Myr ago. These remaining sBHs may be promising monochromatic sources detectable by LISA.

For WDs embedded in AGN disks, the accretion rates are usually higher than that of BHs of similar mass because of much larger star sizes. However, WDs are spun up more efficiently to the shedding limit before reaching the Chandrasekhar limit,
so that it is hard for WDs to reach the Chandrasekhar limit via accretion. On the other hand, it is possible for NSs to grow to the collapse limit via accretion if NS magnetic fields are sufficiently strong to guide the accretion flow outside the NS surface and consequently keep the NS in a slow rotation state while accreting gas. This is a natural way of forming mass-gap-EMRIs in which stellar-mass compact objects are more massive than the heaviest NSs but lighter than $\sim 5M_\odot$.

This paper is organized as follows.
In Section~\ref{sec:model}, we construct a relativistic model of supercritical accretion of BHs, based on which we
study the claimed EM counterpart of GW190521.
In Sections~\ref{sec:BHs} and \ref{sec:NSs}, we apply the supercritical inflow-outflow model
on studying evolution of sBHs, NSs and WDs in AGN disks, respectively. We summarize this paper in Section~\ref{sec:summary}.
Throughout this paper, we use geometrical units $G = c = 1$.

\section{Supercritical inflow-outflow model}\label{sec:model}
In this section, we start with an brief review of the supercritical inflow-outflow model
in Newtonian gravity proposed by Blandford and Begelman \citep{Blandford1999,Blandford2004,Begelman2012}. After that we extend the model to the Kerr spacetime, obtain numerical solutions in the relativistic regime, and finally discuss its application in explaining the EM counterpart of
GW190521.

\subsection{Newtonian model}
In the context of supercritical accretion onto a compact object with mass $m_\bullet$,
part of the inflow gas escapes in the form
of outflow, thus the gas inflow rate $\dot m_{\rm in}(r)$ increases with radius $r$. The dynamics
of steady inflow is governed by the Eulerian equation, the angular momentum equation and the energy equation
\citep{Begelman2012}:
\begin{subequations}\label{eq:dyneq}
\be
 v_r \pder[v_r]{r} = \frac{v_\phi^2}{r}-\frac{m_\bullet}{r^2}-\frac{1}{\rho}\pder[p]{r}\ ,
\ee
\be
\pder{r} (T-\dot m_{\rm in} L) = -(1+\eta_1)L\pder[\dot m_{\rm in}]{r}\ ,
\ee
\be
\pder{r} (T\Omega-\dot m_{\rm in} B) = -(1-\eta_2)B\pder[\dot m_{\rm in}]{r}-2\pi r F^{-}\ ,
\ee
\end{subequations}
where $v_r$ and $v_\phi$ are the velocity in the $r$ and $\phi$ directions, respectively;
$p$ and $\rho$ are the gas pressure and the gas density, respectively;
$L(r)=rv_\phi=r^2\Omega$ is the specific angular momentum;
$B(r)=\frac{v_r^2+v_\phi^2}{2}-\frac{m_\bullet}{r}+\frac{\Gamma}{\Gamma-1}\frac{p}{\rho}$
is the Bernoulli parameter for the inflow material with $\Gamma$ being the adiabatic index, which we take as $\Gamma=4/3$;
$F^-(r) = 2\sigma_{\rm SB}T_{\rm eff}^4$ is the energy flux of radiation cooling
from both surfaces of the disk with $T_{\rm eff}$ being the effective temperature;
$T(r)=-2\pi r^2 \cdot(2H\cdot \nu\rho r \pder[\Omega]{r})$ is the torque that the part of disk with radius $\le r$
exerts on the exterior part, with $\nu$ being the viscosity coefficient relating to the local sound speed $c_s$
and the disk height via $\nu=\frac{2}{3}\alpha c_s H$ following the
commonly used $\alpha$ prescription \citep{Shakura1973}; $\eta_{1,2}$
measures the difference of the specific angular momentum (energy) carried by gas that
turns around from inflow to outflow relative to that of gas still in the inflow at the same radius,
and we take $\eta_1=\eta_2=0$ considering that there should be no discontinuity between the two
\citep[see e.g.,][for the impact of non-zero $\eta_{1,2}$]{Xie2008,Zahra2016}.

The three dynamical equations above are supplemented
by the following contraint equations:
\begin{subequations}\label{eq:coneq}
  \be
  \Sigma = 2\rho H \ ,
  \ee
  \be
-2\pi r \Sigma v_r = \dot m_{\rm in}(r)\ ,
  \ee
  \be
  c_s^2 = \frac{m_\bullet}{r^3} H^2 \ ,
  \ee
  \be
  p = \rho c_s^2
  \ee
  \be
  T_{\rm mid}^4 =\left(\frac{3}{8}\tau + \frac{1}{2}+\frac{1}{4\tau} \right) T_{\rm eff}^4\ ,
  \ee
  \be
  \tau = \frac{\kappa\Sigma}{2}\ ,
  \ee
  \be
    p = p_{\rm rad} + p_{\rm gas}
  \ee
  \be
   p_{\rm rad} =\frac{\tau}{2}\sigma_{\rm SB}T_{\rm eff}^4\ ,
  \ee
  \be
   p_{\rm gas} =\frac{\rho}{m_{\rm p}} k_{\rm B} T_{\rm mid}\ ,
  \ee
\end{subequations}
where $m_{\rm p}$ is the proton mass, $k_{\rm B}$ is the Boltzman constant,
$\Sigma$ is the surface density, $T_{\rm mid}$ is the middle plane temperature, $\tau$ is the optical depth,
$\kappa=\kappa_{\rm sct} +\kappa_{\rm abs}$ is the gas opacity contributed by electron scattering and free-free absorption
with $\kappa_{\rm sct}=0.34\ {\rm cm}^2/{\rm g}$,
and $\kappa_{\rm abs}= 1.7\times 10^{-25} (T_{\rm mid}/{\rm K})^{-7/2} (\rho/({\rm cm}^3{\rm g}^{-1}))
(m_{\rm p}/{\rm g})^{-2} \ {\rm cm}^2/{\rm g}$ \citep{Rybicki1986}.

To close the equations above, one more prescription for the inflow rate $\dot m_{\rm in}(r)$ is needed.
As shown in \citet{Begelman2012}, the condition
\be \dot m_{\rm in}(r) B(r)\approx {\rm const}\ , \ee
must be satisfied to ensure that the Bernoulli parameter of the inflow is bounded from that of the outflow ($B < B_{\rm out}$) everywhere \footnote{$B_{\rm out}$ can be obtained using the energy and angular momentum conservation equations for the outflows, as explained in  \citet{Begelman2012}}.

In inner parts where $p_{\rm rad}>p_{\rm gas}$ and $\kappa_{\rm sct} > \kappa_{\rm abs}$,
there exists a unique self-similar solution to the equations above
with $\dot m_{\rm in}\propto r, \rho\propto r^{-1/2}, p\propto r^{-3/2}$, $H\propto r,
v_r\propto r^{-1/2}, c_s\propto r^{-1/2}, \Omega \propto r^{-3/2}$
and $T_{\rm mid}\propto r^{-3/8}$, $T_{\rm eff}\propto r^{-1/2}$ \citep{Begelman2012,Zahra2016,Zahra2020}.
The self-similar solution in Newtonian gravity is consistent
with numerical simulation results \citep{Yuan2012b,Yuan2012a,Sadowski2014,McKinney2014,Yang2014,Takahashi2018} for
$r\gtrsim 10 m_\bullet$. Similar to ADAFs \citep{Narayan1994,Narayan1997}, the supercritical inflow is radiation inefficient and advection dominated.\footnote{We thank Prof. Feng Yuan for pointing  out this similarity.}
In outer parts where $p_{\rm rad}<p_{\rm gas}$ and/or $\kappa_{\rm sct} < \kappa_{\rm abs}$,
the supercritical inflow becomes even more advection dominated and the radiation cooling is negligible.
In this case, there is no exact self-similar solution, and we can approximately write the solution as:
$\dot m_{\rm in}\propto r, \rho\propto r^{-1/2}, p\propto r^{-3/2}$, $H\propto r,
v_r\propto r^{-1/2}, c_s\propto r^{-1/2}, \Omega \propto r^{-3/2}$
and $T_{\rm mid}\propto r^{-1}$, $T_{\rm eff}\propto r^{-9/8}$ accurate to $\mathcal O(\ln r/r)$.
In outer most parts, the disk becomes unstable due to self-gravity, where the Toomre's stability parameter
\be \label{eq:Toomre}
Q := \frac{c_s\Omega}{\pi \Sigma} \ee
becomes less than unity, and we denote the marginally stable radius where $Q=1$ as $r_Q$.

\subsection{General Relativistic model}
Following \citet{Abramowicz1996}, we extend the above inflow-outflow model to the relativistic regime, with the background spacetime being Kerr.
In the vicinity of the equatorial plane, the Kerr metric is approximately \citep{Page1974}
\be
\begin{aligned}
  ds^2
  &= g_{tt}dt^2 + 2g_{t\phi}dtd\phi +  g_{\phi\phi}d\phi^2 + g_{rr}dr^2+g_{zz}dz^2\ , \\
  &= -\frac{r^2\Delta}{A} dt^2 + \frac{A}{r^2}(d\phi-\omega dt)^2+\frac{r^2}{\Delta} dr^2 + dz^2\ ,
\end{aligned}
\ee
with
\[\Delta = r^2-2m_\bullet r+a^2,\quad A=r^4+r^2a^2+2m_\bullet ra^2, \quad \omega = \frac{2m_\bullet ar}{A}\ .\]
Assuming the four-velocity of matter has three non-vanishing components
\be
 u^\mu = (u^t, u^r, 0, u^\phi)\ ,
\ee
we define the angular velocity $\Omega$ relative to the stationary observer and the angular velocity
$\tilde\Omega$ relative to the local inertial observer by
\be
\Omega = \frac{u^\phi}{u^t} , \quad \tilde \Omega = \Omega -\omega\ .
\ee
In particular, the angular velocity of the corotating and counterrotating Keplerian orbits are
\be
\Omega_{\pm}=\pm\frac{m_\bullet^{1/2}}{r^{3/2}\pm am_\bullet^{1/2}}\ .
\ee
In accordance with \citet{Abramowicz1996}, we define a rescaled radial velocity $V$ by
\be
\frac{V}{\sqrt{1-V^2}} = u^r \sqrt{g_{rr}}\ ,
\ee
and the Lorentz factors of the total motion $\gamma$ and of the $\phi$-direction motion  $\gamma_\phi$
\be
\gamma = \frac{1}{\sqrt{1-\tilde\Omega^2\tilde R^2}}\frac{1}{\sqrt{1-V^2}},
\quad \gamma_\phi = \frac{1}{\sqrt{1-\tilde\Omega^2\tilde R^2} }\ ,
\ee
with $\tilde R^2 = \frac{A^2}{r^4\Delta}$.
As a result, the specific angular momentum and the specific energy are given by
\be
L = \gamma \frac{A^{3/2}}{r^3\Delta^{1/2}} \tilde \Omega\ ,\quad
E = -\frac{g_{tt}+g_{t\phi}\Omega}{g_{t\phi}+g_{\phi\phi}\Omega} L \ .
\ee

With the definitions above, the three dynamical equations in Eq.~(\ref{eq:dyneq}) written in the relativistic form become
\begin{subequations}
  \be
   \frac{1}{1-V^2}\frac{V^2}{c_s^2} \pder[\ln V]{\ln r}
    =\frac{\mathcal A}{c_s^2} -\pder[\ln\rho ]{\ln r} -2\pder[\ln c_s]{\ln r}\ ,
  \ee
  \be
  \frac{1}{2\pi}\left[  (\dot m_{\rm in}L- \dot m_{\rm in,0}L_0)-\int_{r_0}^r L \frac{d\dot m_{\rm in}}{dr} dr\right]
  = -\nu\frac{\Sigma A^{3/2}\Delta^{1/2}\gamma^3}{r^4} \pder[\Omega]{r}\ ,
  \ee
  \be
  \dot m_{\rm in} \left( \frac{1}{\Gamma-1}\pder[c_s^2]{\ln r}-c_s^2\pder[\ln \rho]{\ln r} \right)
  = -2\pi r^2 \left(F^+-F^-\right)\ ,
  \ee
\end{subequations}
with
\[
\mathcal A = -\frac{m_\bullet A}{r^3\Delta \Omega_+\Omega_-}
\frac{(\Omega-\Omega_+)(\Omega-\Omega_-)}{1-\tilde\Omega^2\tilde R^2}\ , \]
where $\dot m_{\rm in,0}$ is the gas inflow rate on the horizon,
$L_0$ is the specific angular momentum of gas on the horizon,
the viscosity heating rate is given by $F^+ = \nu\Sigma \gamma^4 \left(\frac{A}{r^3}\pder[\Omega]{r}\right)^2$
and the radiation cooling rate is $F^- = 2\sigma_{\rm SB}T_{\rm eff}^4$.

Similar to Eq.~(\ref{eq:coneq}), the constraint equations in the relativistic form are
\begin{subequations}
  \be
  \Sigma = 2\rho H
  \ee
  \be
  -2\pi  \Sigma \sqrt{\Delta} \frac{V}{\sqrt{1-V^2}} = \dot m_{\rm in}(r)
  \ee
  \be
  c_s =  \gamma H  \sqrt{\frac{m_\bullet}{r^3}
  \frac{(r^2+a^2)^2+2\Delta a^2}{(r^2+a^2)^2-\Delta a^2}} :=\gamma H\Omega_\perp\ ,
  \ee
  \be
  p = \rho c_s^2
  \ee
  \be
  T_{\rm mid}^4 =\left(\frac{3}{8}\tau + \frac{1}{2}+\frac{1}{4\tau} \right) T_{\rm eff}^4\ ,
  \ee
  \be
  \tau = \frac{\kappa\Sigma}{2}\ ,
  \ee
  \be
  p_{\rm rad} =\frac{\tau}{2}\sigma_{\rm SB}T_{\rm eff}^4\ ,
  \ee
\end{subequations}
where we have used the approximation $p\approx p_{\rm rad}$ because we will solve the relativistic equations
only in radiation-dominated parts.

In the relativistic regime, the solution differs from the Newtonian  self-similar solution as there is an additional characteristic radius, sonic point $r_{\rm cs}$, where the inflow velocity surpasses the local sound speed  (see next subsection for explicite definition of  $r_{\rm cs}$). Based on the obervation that the outflow dies down around the characteristic radius \citep{Yuan2012b,Yuan2012a,McKinney2014,Sadowski2014,Yang2014,Takahashi2018},
we formulate the prescription of inflow rate as
\be\label{eq:mdot}
\begin{aligned}
  \dot m_{\rm in}(r) B(r) &= {\rm const}\quad (r \geq r_{\rm cs})\ ,\\
  \dot m_{\rm in}(r)  &= {\rm const}\quad (r < r_{\rm cs})\ .
\end{aligned}
\ee
with the relativistic Bernoulli parameter being $B=E-1+\frac{\Gamma}{\Gamma-1}Ec_s^2$.

\subsection{Numerical solutions to the relativistic model}

In practice, we solve the relativistic disk structure for $r\leq 10^3 m_\bullet$ and
use the Newtonian solution  for $r\in (10^3 m_\bullet, r_Q)$,
where $r_Q$ is the marginally stable radius [see Eq.~(\ref{eq:Toomre})].
In order to numerically solve the relativistic disk structure,
we reorganize the dynamical equations (with the aid of the constraint equations) as
\be\label{eq:dOmega}
\pder[\Omega]{r}
= V \frac{r^4\Omega_\perp}{\frac{2}{3}\alpha\gamma\gamma_\phi c_s^2 A^{3/2} }
\left(L-\frac{\dot m_{\rm in,0}}{\dot m_{\rm in}}L_0 -\frac{1}{\dot m_{\rm in}} \int_{r_0}^r L \frac{d\dot m_{\rm in}}{dr} dr\right)\ ,
\ee
\be\label{eq:dV}
\begin{aligned}
  \pder[\ln |V|]{\ln r}
  =\frac{ \frac{\mathcal A}{c_s^2}
  -\frac{2\Gamma}{\Gamma+1} \left(\pder[\ln ...]{\ln r}\right)
   +\frac{\Gamma-1}{\Gamma+1}\delta F}
   {\left(\frac{1}{1-V^2}\frac{V^2}{c_s^2}-\frac{2\Gamma}{\Gamma+1}  \right)}
   \ ,
\end{aligned}
\ee
\be\label{eq:dcs}
\begin{aligned}
   \pder[\ln c_s]{\ln r}=
   \frac{-\frac{\mathcal A}{c_s^2}
  +\left(\frac{1}{1-V^2}\frac{V^2}{c_s^2} \right)\left(\pder[\ln ...]{\ln r} \right)
  -\left(\frac{1}{1-V^2}\frac{V^2}{c_s^2}-1\right)\delta F}
  {\left(\frac{1}{1-V^2}\frac{V^2}{c_s^2}-\frac{2\Gamma}{\Gamma+1}  \right)\frac{\Gamma+1}{\Gamma-1}}\ ,
\end{aligned}
\ee
with
\[\delta F = \frac{2\pi r^2}{\dot m_{\rm in} c_s^2}(F^+-F^-)\ ,\]
\[ \left(\pder[\ln ...]{\ln r}\right) = \left(\pder[\ln\dot m]{\ln r}+\pder[\ln\gamma_\phi]{\ln r}
  +\pder[\ln \Omega_\perp]{\ln r}
  -\pder[\ln \sqrt{\Delta}]{\ln r}\right)\ .
\]
Eq.~(\ref{eq:dV}) is singular at the sonic point $r_{\rm cs}$ where the denominator of the right-hand side vanishes, while a physical solution must also have a vanishing  numerator at the same radius $r=r_{\rm cs}$.
For a given inflow rate $\dot m_{\rm in}(r)$ and a sonic point $r_{\rm cs}$,
solving the three equations above is an eigenvalue problem
with the to-be-determined eigenvalue $L_0$.
To close the eigenvalue problem with three independent variables and one eigenvalue, we need to specify
4 boundary conditions, which we choose according to \cite{Narayan1997}: vanishing numerator and denominator of
Eq.~(\ref{eq:dV}) at the sonic point $r=r_{\rm cs}$,  matching $c_s$ and $\Omega$ with their counterparts in the Newtonian self-similar solution at  $r=10^3 m_\bullet$. Starting with  a guess solution,
we solve the above eigenvalue problem using the relaxation method \citep{Press2002},
and update the inflow rate $\dot m(r)$ according to the prescription (\ref{eq:mdot})
every a few steps of relaxation.
We finally obtain the converged solution $\{V(r), c_s(r), \Omega(r), L_0\}$ in the range
$r\in (r_{\rm cs}, 10^3 m_\bullet)$.
With the eigenvalue $L_0$, we again solve the three dynamical equations from the horizon $r_{\rm H}$ to the
sonic point $r_{\rm cs}$
using the same relaxation method with 3 boundary conditions: vanishing numerator and denominator of
Eq.~(\ref{eq:dV}) at $r=r_{\rm cs}$, and $L=L_0$ at $r=r_{\rm H}$.
For an initial guess of $r_{\rm cs}$, the two solution pieces generally do not match at $r=r_{\rm cs}$.
Therefore, we need to adjust the location $r_{\rm cs}$
of the sonic point and repeat the calculation above until a global solution is found.

\begin{figure}
  \includegraphics[scale=0.8]{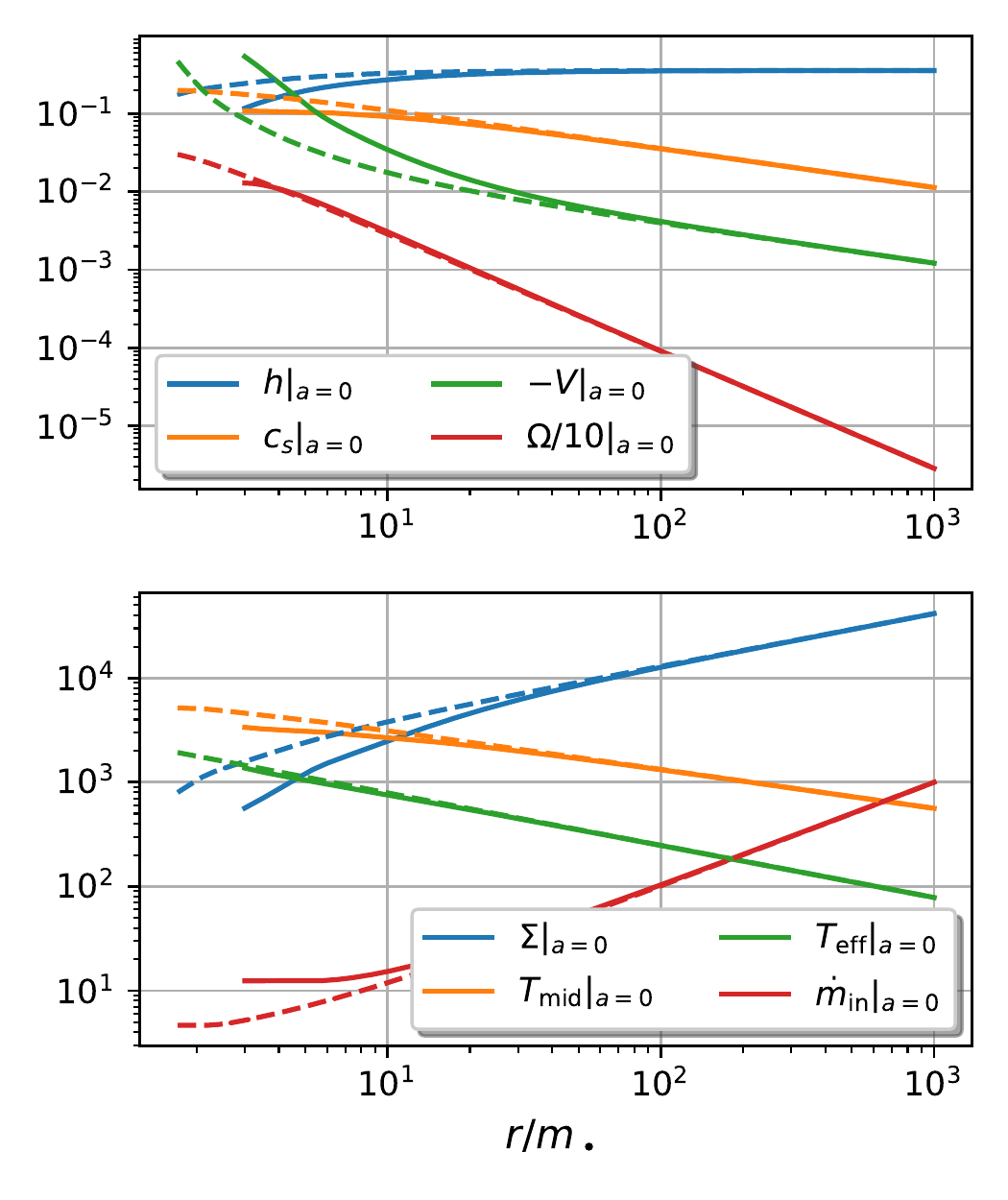}
  \caption{Numerical solutions to the supercritical inflow-outflow model.
  Solid/dashed lines are   disk aspect ratio $h$, radial velocity $V$, sound speed $c_s$, angular velocity $\Omega\ [m_\bullet^{-1}]$, surface density $\Sigma\ [{\rm g}/{\rm cm}^2]$,
 middle plane temperature $T_{\rm mid}\ [\rm eV]$,
  effective temperature $T_{\rm eff}\ [\rm eV]$, and gas inflow rate $\dot m_{\rm in}\ [\medd]$
  for a BH with mass $m_\bullet=10 M_\odot$ and spin $a=0/0.9$.}
  \label{fig:sd}
\end{figure}

In Fig.~\ref{fig:sd}, we show two sample solutions with $m_\bullet=10 M_\odot$, $a=0$ or $0.9$, $\dot m_{\rm in} = 10^3 \medd$ at $r=10^3 m_\bullet$ and $\alpha=0.1$, where we have defined the Eddington accretion rate as
$\medd = 10 L_\bullet^{\rm Edd}$, with the Eddington luminosity $L_\bullet^{\rm Edd}=1.26\times 10^{38} (m_\bullet/M_\odot) $ erg/s. Similar to  slim disks \citep{Sadowski2009} and ADAFs \citep{Abramowicz1996, Narayan1997}, we find the sonic point location $r_{\rm cs}$ of the inflow-outflow model is also close to the innermost stable circular orbit (ISCO) $r_{\rm isco}$ regardless of the BH spin.  As expected, both sample solutions approach the Newtonian self-similar
solution for large $r$, and become increasingly different close to the BH. Compared to simulations of General Relativistic magnetohydrodynamics (GRMHD) \citep[e.g.,][]{Sadowski2014}, we find the simple steady inflow-outflow model reproduces the dependence of the disk structure on the BH spin $a$: as $a$ increases, both the surface density $\Sigma(r)$ and  the  luminosity (or equivalently the higher effective temperature $T_{\rm eff}(r)$) grows, whereas the  radial velocity $-V(r)$ and the specific angular momentum  $L(r)$ decreases. The prescription for the inflow rate $\dot m_{\rm in}(r)$ (Eq.~\ref{eq:mdot}) is well consistent with simulation results for large $r$, while the transition radius varies in
different simulations \citep{Yuan2012b,Yuan2012a,McKinney2014,Sadowski2014,Yang2014,Takahashi2018}, and the choice of  the sonic radius $r_{\rm cs}$ as the transition radius also reproduces its spin dependence found in these simulations.

\begin{figure}
  \includegraphics[scale=0.8]{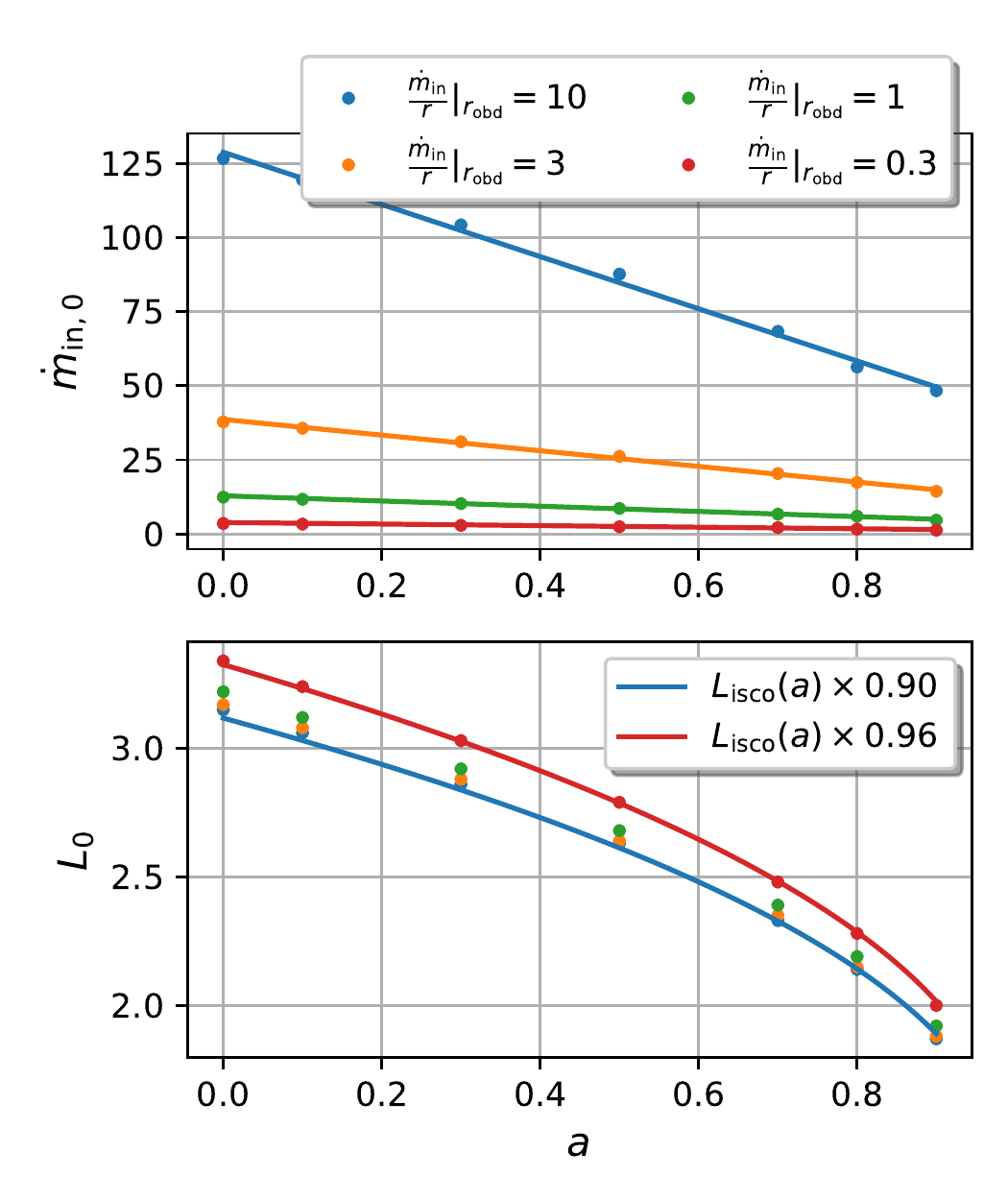}
  \caption{Upper panel: the BH accretion rate $\dot m_{\rm in,0}\ [\dot m_{\bullet}^{\rm Edd}]$ dependence on the BH spin $a$
  and the inflow rate at the outer boundary $\dot m_{\rm in}/r\ [\dot m_{\bullet}^{\rm Edd}/m_\bullet]$, where the discrete dots
  are numerical results and the solid lines are linear fitting of Eq.~(\ref{eq:mdot0_fit}).
  Lower panel: the dependence of the specific angular momentum $L_0$ of gas falling into the horizon on the BH spin,
   where the two solid lines are  $0.9L_{\rm isco}(a)$ and $0.96L_{\rm isco}(a)$, respectively.
  }
  \label{fig:mdot0_L0}
\end{figure}

We numerically solve the disk structure for different BH masses, spins, $\alpha$ values and inflow rates $\dot m_{\rm in}$ at the outer boundary $r_{\rm obd}=10^3 m_\bullet$. We find the dependence of the BH accretion rate $\dot m_{\rm in,0}$
on the BH spin $a$ and the inflow rate at the outer boundary can be well fitted by
\be
\label{eq:mdot0_fit}
\frac{\dot m_{\rm in,0}}{\dot m_{\bullet}^{\rm Edd}} = (12.87-8.80a)\times
\frac{\dot m_{\rm in}(r_{\rm obd})/r_{\rm obd} }{\medd/m_\bullet}\ ,
\ee
and the specific angular momentum $L_0(a)$ and the specific energy $E_0(a)$ of gas falling into the horizon
are approximated by $L_{\rm isco}(a)$ and $E_{\rm isco}(a)$ accurate to $10\%$ or better, respectively, where
$L_{\rm isco}(a)$ and $E_{\rm isco}(a)$ are the specific angular momentum and specific energy of particles on the ISCO $r_{\rm isco}(a)$ \citep{Bardeen1972}, i.e.,
\be\label{eq:EL_fit}
\begin{aligned}
  E_0(a)\approx E_{\rm isco}(a) &= \frac{\risco^{3/2}-2\risco^{1/2}+a}{\risco^{3/4}\left(\risco^{3/2}-3\risco^{1/2}+2a\right)}\ , \\
  L_0(a)\approx L_{\rm isco}(a) &= \frac{\risco^{2}-2a\risco^{1/2}+a^2}{\risco^{3/4}\left(\risco^{3/2}-3\risco^{1/2}+2a\right)}\ .
\end{aligned}
\ee
Note that both Eq.~(\ref{eq:mdot0_fit}) and Eqs.~(\ref{eq:EL_fit}) are independent of the BH mass $m_\bullet$ and the $\alpha$ parameter. In Fig.~\ref{fig:mdot0_L0}, we show the numerical results of the BH accretion rate $\dot m_{\rm in,0}$ and the specific angular momentum $L_0$ of gas falling into the horizon for $m_\bullet=10 M_\odot$ and $\alpha=0.1$, along with the comparison to Eqs.~(\ref{eq:mdot0_fit},\ref{eq:EL_fit}).

\subsection{Radiation efficiency of supercritical accretion and the EM counterpart of GW190521}

As noted in previous studies \citep{McKinney2014,Sadowski2014,Yang2014},
the supercritical accretion is radiation inefficient in terms of radiation efficiency
$\eta_{\rm rad}:= L_{\rm rad}/\dot m_{\rm in}(r_{\rm obd})=\int 2\pi r E F^- d r/\dot m_{\rm in}(r_{\rm obd})$.
For the sample solution in Fig.~\ref{fig:sd} with spin $a=0$, the disk luminosity $L_{\rm rad}\approx 4.5 L_\bullet^{\rm Edd}$ wheras the mass inflow rate at the outer boundary is $10^3 \dot m_{\bullet}^{\rm Edd}=10^4 L_\bullet^{\rm Edd}$, therefore
the radiation efficiency is $\eta_{\rm rad} \approx 0.045\%$. For the other solution with spin $a=0.9$, the disk luminosity is $L_{\rm rad}\approx 5.2 L_\bullet^{\rm Edd}$ and the radiation efficiency is $\eta_{\rm rad} \approx 0.052\%$.

Another interesting feature of the supercritical inflow-outflow model is that the disk luminosity
is nearly independent of the inflow rate $\mdoto$ at the outer boundary,  because both
$T_{\rm mid}^4$ and the optical depth $\tau$ are proportional to $\mdoto$, so that the effective temperature
$T_{\rm eff}$ turns out to be nearly independent of $\mdoto$. In other words, the radiation cooling rate does not change
because the disk becomes hotter ($T_{\rm mid}$) and denser ($\tau$) at the same time.
As a result, the supercritical accretion becomes more advection dominated and more radiation inefficient
for higher $\mdoto$. The radiation efficiency for $\mdoto/r_{\rm obd} >  \medd/m_\bullet$ can be fitted as
\be\label{eq:eta_io}
\eta_{\rm rad} \approx 0.5 \left(\frac{\mdoto}{\medd}\right)^{-1}\ ,
\ee
with mild dependence on the BH spin.

For comparison, the supercritical accretion of BHs without considering outflow has also been investigated by \citet{Beloborodov1998},
where the supercritical accretion is also found to be highly advection dominated.
For accretion rate $\dot m_\bullet > 10 \medd$, the radiation efficiency can be fitted as
\be\label{eq:eta_in}
\eta_{\rm rad}\approx 0.18\left(\frac{\dot m_\bullet}{\medd}\right)^{-0.7}\ ,
\ee
with mild dependence on the BH spin. Therefore, the radiation efficiency of supercritical accretion
(with or without outflow) is low and decreases with the accretion rate.

GW190521 is a merger of two BHs with masses of $\sim 85 M_\odot$ and $\sim 66 M_\odot$ \citep{GW190521},
and an optical flare with luminosity $\sim 10^{45}$ erg/s detected by  Zwicky Transient Facility has been proposed as the EM counterpart of the merger \citep{Graham2020}. In \citet{Graham2020}, the optical flare is interpreted as
the emission from the Bondi-Hoyle-Lyttleton (BHL) accretion of the remnant BH in an AGN disk with accretion rate and luminosity
\be
\begin{aligned}
  \dot M_{\rm BHL} &= 2.8\times10^{25}\ {\rm g/s} \left(\frac{M_{\rm BBH}}{100 M_\odot}\right)^2
  \left(\frac{v_{\rm rel}}{100 \ {\rm km/s}}\right)^{-3} \\
  &\times \left(\frac{\tilde\rho}{10^{-10}\ {\rm g/cm^3}}\right) \approx 2\times10^5 \dot M_\bullet^{\rm Edd}\ ,\\
  L_{\rm rad} &\approx 2.5\times 10^{45} \  {\rm erg/s} \left(\frac{\eta_{\rm rad}}{0.1}\right) \approx 2\times10^5 L_\bullet^{\rm Edd}\ ,
\end{aligned}
\ee
where $M_{\rm BBH}$ is the mass of the remnant BH,  $\tilde\rho$ is the local gas density in the AGN disk, $v_{\rm rel}$ is the relative velocity of the remnant BH w.r.t the local gas.
However the assumed radiation efficiency $\eta_{\rm rad}=0.1$ largely overestimates the true value in the case of supercritical accretion (with or without outflow) with inflow rate as high as $2\times10^5 \dot M_\bullet^{\rm Edd}$.
In the supercritical inflow-outflow model, there is a upper limit of the disk luminosity $\lesssim 10 L_\bullet^{\rm Edd}$
as explained above, so that the luminosity will be orders of magnitude less than the level $\sim 2\times 10^5 L_\bullet^{\rm Edd}$ \citep[see also][]{McKinney2014,Sadowski2014,Yang2014}.
On the other hand, even in the supercritical accretion model without considering outflow, the radiation efficiency is about
$\eta_{\rm rad}\approx 3.5\times 10^{-5}$ for $\dot M_{\rm BHL}=2\times 10^5 \dot M_\bullet^{\rm Edd}$.
Therefore, the claimed optical flare seems unlikely to be explained by the emission from the supeircritical accretion flow onto  the remnant BH.

In fact,  in the inflow-outflow model, most of the outflows are generated at radius comparable to $r_{\rm obd}$, as infered by the radius dependence of $\dot{m}_{\rm in}$. As a result, these gas materials have not fully released the gravitational potential energy, as compared with the small fraction of gas that accretes onto the BH. In other words, the energy reservoir from the accretion flow roughly scales as $\dot{m}_{\rm in}(r_{\rm obd})/r_{\rm obd}$, which is insufficient to power the proposed optical flare in \citet{Graham2020}.

\section{Evolution of sBHs in an AGN disk}\label{sec:BHs}
For a stellar cluster around an accreting MBH, some of stars and sBHs inside the cluster are captured by the disk, so that the total population settles into a cluster component and a disk component as interacting with the disk \citep{Pan2021}.
SBHs embedded in the AGN disk are expected to spin up and grow in mass via accretion as migrating toward the MBH.
In this section, we will first introduce the commonly used AGN disk models, summarize the interactions of
stars and sBHs with the AGN disk, then explore the mass and spin evolution of the captured sBHs inside the disk and
finally calculate their spatial distribution.

\subsection{AGN disk models}\label{subsec:AGN_disks}
The structure of AGN disks has not been fully understood due to two main uncertainties:  the mechanism
of disk angular momentum transport and  the mechanism of disk heating in outer parts
where the local viscosity heating is not sufficient. In this work, we consider three commonly used
AGN disk models: $\alpha/\beta$ disk \citep{Sirko2003} and TQM disk \citep{Thompson2005}.
In the first two models, the disk momentum transport mechanism is characterized by the canonical disk viscosity,
where the viscous stress is parameterized as $\alpha \tilde p$/$\alpha \tilde p_{\rm gas}$ in the $\alpha/\beta$  disk model,
with $\alpha \lesssim 1$ being a phenomenological parameter, $\tilde p$ being the total pressure and $\tilde p_{\rm gas}$ being the gas pressure. In the outer parts of the disk, certain external heating process is  assumed to
keep the disk to be marginally stable against the disk self-gravity.
In the TQM disk model, the disk angular momentum is assumed to be carried away by more efficient global
torques and the gas inflow velocity is parameterized as a constant fraction of local sound speed
i.e., $-\tilde v_r = X \tilde c_s$;  in outer parts of the disk, star formation in the disk is self-consistently
implemented to heat the disk and maintain its stability against the disk self-gravity.\footnote{Very Recently,
\citet{Gilbaum2021} proposed a feedback-dominated accretion flow model in which outer parts of an AGN disk is heated by
EM emissions from accreting sBHs embedded in the AGN disk.}
In this work, we use tilde variables to distinguish AGN disk variables from variables of small disks around stellar-mass objects.

\begin{figure*}
  \includegraphics[scale=0.6]{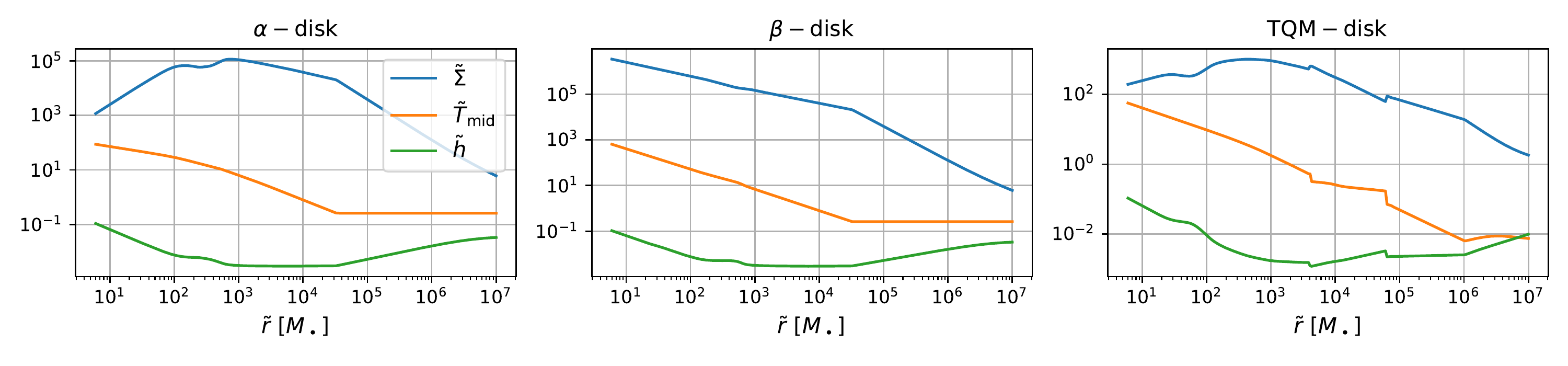}
  \caption{Three commonly used AGN disk models for a MBH with $M_\bullet=4\times 10^6 M_\odot, \dot M_\bullet = 0.1 \Medd$, where $\alpha=0.1$ for the $\alpha/\beta$ disk, and $X=0.1$ for the TQM disk.
  The three disk variables shown are the disk surface density $\tilde \Sigma \ [\rm g/cm^2]$, the disk middle-plane temperature $\tilde T_{\rm mid}\ [\rm eV]$ and the disk aspect ratio $\tilde h$.}
  \label{fig:disks}
\end{figure*}

In Fig.~\ref{fig:disks}, we show the solutions in the three AGN disk models for a MBH with $M_\bullet=4\times10^6 M_\odot, \dot M_\bullet = 0.1 \Medd$, where we have used $\alpha=0.1$ for the $\alpha/\beta$  disks and $X=0.1$ for the TQM disk \citep[see][for details]{Pan2021}.
From Fig.~\ref{fig:disks}, the $\alpha$ disk and the $\beta$ disk become different from each other only at small $\tilde r$
where the gas pressure becomes subdominant. This means that the viscous tress and the gas inflow velocity  in the  $\beta$ disk become
smaller than those in the $\alpha$ disk while the surface density becomes larger. One major difference of the solution of  the TQM disk is that the gas inflow velocity is much higher than the value in the  other two disks, therefore the surface density is much lower.

The $\alpha$-viscosity prescription is known as a good approximation to the turbulence viscosity driven by
magnetorotational instability in inner parts of accretion disks where the gas is fully
ionized \citep{Balbus1991,Balbus1998,Martin2019}. Therefore we expect $\alpha/\beta$ disk models should be
a closer description to inner parts ($\tilde r\lesssim 10^5 M_\bullet$ for the disks in Fig.~\ref{fig:disks})
of AGN disks in nature, while it is not clear which model works better in outer parts.

\subsection{Density waves, head wind and supercritical accretion}\label{subsec:interactions}
Periodic motion of a sBH excites density waves that drive the  inward migration of sBHs, circularize the planet’s orbit and drive the sBH orbit toward the disk plane \citep{Goldreich1979,Goldreich1980,Ward1989,Tanaka2002,Tanaka2004}. Here we focus on the inward migration of a sBH in the AGN disk.
Previous analytic studies together with numerical simulations show that the specific migration
torque arising from density waves is given by \citep{Tanaka2002,Tanaka2004}
\be\label{eq:JI}
\dotJI = C_{\rm I}\frac{m_\bullet}{M} \frac{\tilde \Sigma}{M }\frac{\tilde r^4\Omega_\bullet^2}{\tilde h^2}\ ,
\ee
where  $M=M(<r)$ is the total mass within radius $r$, $C_{\rm I} = -0.85 + d\ln \tilde\Sigma/d\ln \tilde r + 0.9\ d\ln \tilde T_{\rm mid}/d\ln \tilde r$ and $\Omega_\bullet$ is the angular velocity of the sBH around the MBH \citep{Paardekooper2010}.

Gas inside the influence sphere of the sBH tends to flow towards it.
Due to the differential rotation of the disk, the nearly radial inflow at large separation, which generally carries nonzero angular momentum,
becomes circularized at smaller seperation. A disk consequently forms at the circularization radius, which serves as the outer
boundary $r_{\rm obd}$ of the disk model. As described by the inflow-outflow model in the previous section, a large fraction
of gas supplied at the outer boundary finally escapes in the form of outflow, and only the remaining part accretes onto
the sBH. Because of the circularization process, it is reasonable to expect that the outflow carries minimal amount of net momentum w.r.t. the sBH. As a result, the ``head wind" w.r.t the sBH are captured at places where the sBH gravity becomes important, and the momentum carried by the wind eventually transfers to the sBH.
The specific torque exerted on the sBH from the wind is written as
\be\label{eq:Jwind}
\dot J_{\rm wind}  = - \frac{\tilde r \delta v_\phi \dot m_{\rm wind}}{m_\bullet}\ ,
\ee
where $\delta v_\phi$ is the relative bulk velocity in the $\phi$ direction.
The head wind strength $\dot m_{\rm wind}$ (which determines the gas inflow rate $\dot m_{\rm in}(r_{\rm obd})$
of the sBH accretion disk at the outer boundary $r_{\rm obd}$) can be estimated
according to the BHL rate $\dot m_{\rm BHL}$
with corrections accounting for the limit of MBH gravity in the radial direction
and the finite size of the AGN disk in the vertical direction \citep{Kocsis2011}, i.e.,
\be\label{eq:mdot_obd}
\dot m_{\rm wind} = \dot m_{\rm BHL} \times {\rm min}\left\{1, \frac{\tilde H}{r_{\rm BHL}}\right\}\times
{\rm min}\left\{1, \frac{r_{\rm Hill}}{r_{\rm BHL}}\right\}\ ,
\ee
and the circularization radius $r_{\rm cir}$ is estimated from the angular momentum conservation of infalling gas, i.e.,
\be\label{eq:r_obd}
\sqrt{r_{\rm cir} m_\bullet} = v_{\rm rel}(r_{\rm rel}) r_{\rm rel}\ ,
\ee
with $r_{\rm rel}:={\rm min}\{r_{\rm BHL}, r_{\rm Hill}\}$, $\tilde H$ being the height of the AGN disk and $r_{\rm Hill}=(m_\bullet/3M)^{1/3}\tilde r$ being the Hill radius of the sBH.
The Bondi accretion rate and the Bondi radius are  known as
\be\label{eq:mdot_Bondi}
\frac{\dot m_{\rm BHL}}{m_\bullet} = \frac{4\pi\tilde \rho m_\bullet}{(v_{\rm rel}^2 +\tilde c_s^2)^{3/2}}\ ,
\ee
and $r_{\rm BHL} = m_\bullet/(v_{\rm rel}^2 +\tilde c_s^2)$,
where $v_{\rm rel}$ is the relative velocity between the sBH and the local gas,
$v_{\rm rel}^2 = (\delta v_\phi + \delta v_{\rm dr})^2 + \delta v_r^2$,
with  the relative bulk velocity $\delta v_r$ in the  $r$ direction,
and   the relative velocity $\delta v_{\rm dr}$ coming from the differential rotation
of the AGN disk
\citep{Kocsis2011}:
\begin{subequations}\label{eq:v_rel}
  \be
  \delta v_\phi = \frac{3-\tilde\gamma}{2} \tilde h \tilde c_s\ ,
  \ee
  \be
  \delta v_r = |\tilde v_{r,\rm gas}-v_{r,\bullet}|
  = \left|-\frac{\dot M_\bullet}{2\pi \tilde r\tilde \Sigma}-\frac{\dot J}{dJ/d\tilde r}\right|\ ,
  \ee
  \be
  \delta v_{\rm dr} = \frac{3}{2}\frac{r_{\rm rel}}{\tilde r} \tilde h^{-1} \tilde c_s\, .
  \ee
\end{subequations}
Here $\tilde \gamma$ is defined as $d\ln\tilde\rho/d\ln \tilde r$,  $\dot J$ is the specific torque exerted on the sBH
due to sBH-disk interactions and the GW emission, i.e., $\dot J = \dot J_{\rm mig, I,II} + \dot J_{\rm wind} + \dot J_{\rm gw}$,
where
\be\label{eq:Jgw}
\dot J_{\rm gw} = -\frac{32}{5} \frac{m_\bullet}{M}\left(\frac{M}{\tilde r}\right)^{7/2}\ ,
\ee
$J=\tilde r^2\Omega_\bullet$, $\dot J_{\rm mig,I,II}= \dot J_{\rm mig,I}$ or $\dot J_{\rm mig,II}$,
$ \dot J_{\rm wind}= \dot J_{\rm wind}$ or $0$ depending on whether a gap is opened.
For later convenience, we define the migration timescale of the sBH as
\be\label{eq:tmig}
t_{\rm mig} := \frac{\tilde r}{|\dot{\tilde r}|}
= \frac{1}{2}\frac{J}{|\dot J|}\ .
\ee

\begin{figure*}
  \includegraphics[scale=0.6]{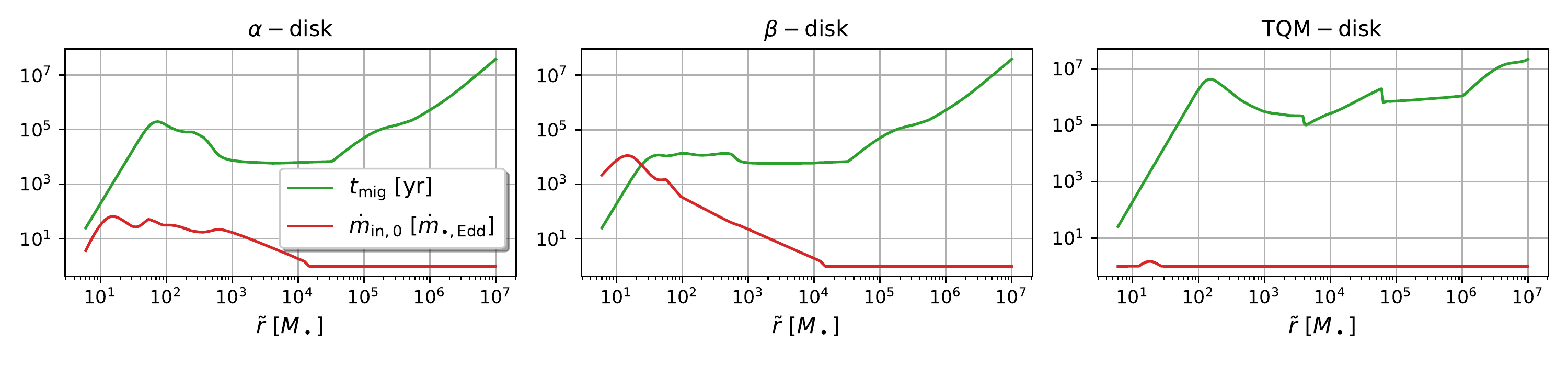}
  \caption{Migration timescales $t_{\rm mig}$ [Eq.~(\ref{eq:tmig})]
  and accretion rates $\dot m_{\rm in,0}$  [Eq.~(\ref{eq:mdot0})]  of sBHs with mass $m_\bullet=10 M_\odot$ and spin $a=0$ embedded in fiducial AGN disks (Fig.~\ref{fig:disks}).}
  \label{fig:mdot_tmig}
\end{figure*}

For a sBH embbeding in an AGN disk, the wind strength $\dot m_{\rm wind}$ and the circularization radius $r_{\rm cir}$
of inflowing gas are obtained from Eqs.~(\ref{eq:mdot_obd}-\ref{eq:Jgw}). If the local disk around the sBH
is stable against self-gravity at $r_{\rm cir}$, i.e., $r_{\rm cir} < r_Q$, the outer boundary conditions
of the local disk are naturally
\be
r_{\rm obd} = r_{\rm cir}, \quad \dot m_{\rm in}(r_{\rm obd}) = \dot m_{\rm wind}\ .
\ee
On the other hand, if the local disk is unstable at $r_{\rm cir}$, i.e., $r_{\rm cir} > r_Q$,
with gas feeding rate $\dot m_{\rm wind}$, the gas inflow $\dot m_{\rm in}(r_{\rm obd})$
is consequently suppressed by the self-gravity instability
until a larger marginally stable radius $r_Q\approx r_{\rm cir}$ (which increases with decreasing accretion rate)
is reached. Thus the boundary conditions are formulated as
\be
r_{\rm obd} = r_{\rm cir}, \quad r_Q(\dot m_{\rm in}(r_{\rm obd})) = r_{\rm cir}\ ,
\ee
where  $\dot m_{\rm in}(r_{\rm obd})<\dot m_{\rm wind}$.

\begin{figure*}
  \includegraphics[scale=0.6]{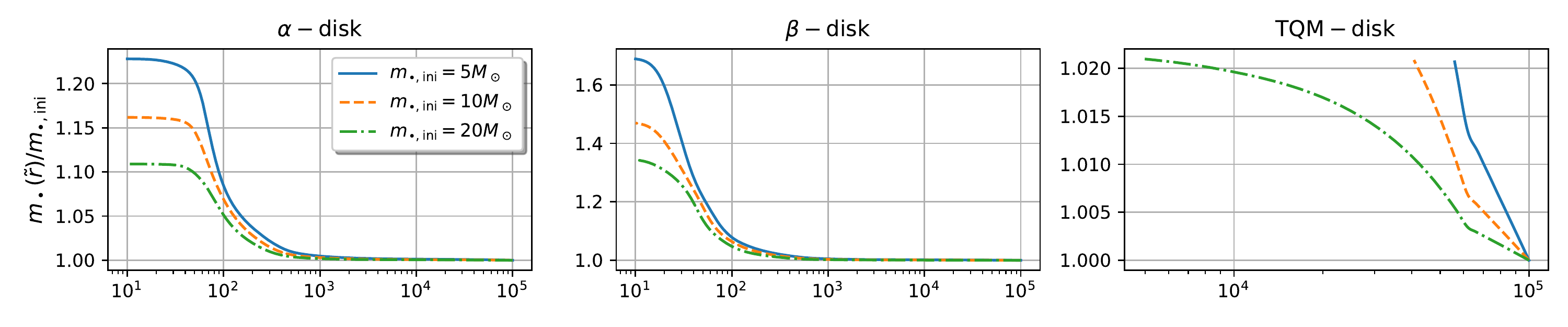}
  \includegraphics[scale=0.6]{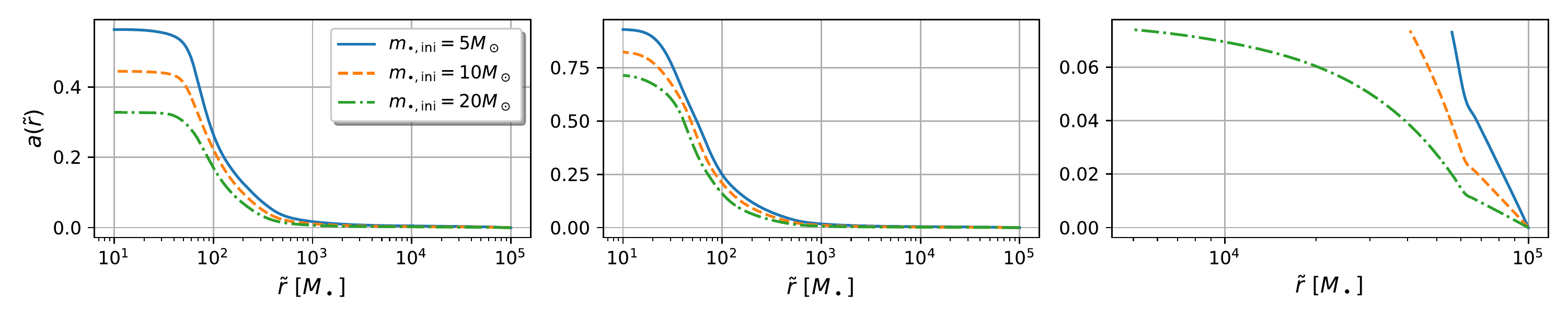}
  \caption{The mass and spin evolution of sBHs for a duration
  of $10^6$ yr with initial mass $m_{\bullet, \rm ini}=5/10/20 M_\odot$,
  initial spin $a_{\rm ini}=0$ and initial location $\tilde r_{\rm ini} = 10^5 M_\bullet$. }
  \label{fig:mass_spin}
\end{figure*}

With the above boundary conditions, the disk structure can be modelled and numerically solved
as shown in the previous section. As a result, we obtain the sBH accretion rate as
\be\label{eq:mdot0}
\frac{\dot m_{\rm in,0}}{\dot m_{\bullet}^{\rm Edd}}=
\begin{cases}
   \dot m_{\rm in}(r_{\rm obd})/\medd\ , \hfill (m_{\rm in}(r_{\rm obd}) < \medd) \\
  {\rm max}\left\{ (12.87-8.80a)
  \frac{\dot m_{\rm in}(r_{\rm obd})/r_{\rm obd} }{\dot m_{\bullet,\rm Edd}/m_\bullet}, 1\right\}\ , \hfill ({\rm otherwise})
\end{cases}
\ee
following Eq.~(\ref{eq:mdot0_fit}) and considering the possiblity that the supercritical
inflow-outflow model may not hold where the inflow rate $\dot m_{\rm in}$ falls below the Eddington rate, so that the disk decription is switched to the one without outflow at subcritical rates.
In the next subsection (Fig.~\ref{fig:mass_spin}),
we will see this prescription of minimum sBH accretion rate has little impact
on the mass and spin evolution of sBHs embedded in an AGN disk.

In Fig.~\ref{fig:mdot_tmig}, we show the sBH accretion rates $\dot m_{\rm in,0}$ and  migration timescales $t_{\rm mig}$
of sBHs of mass $10 M_\odot$ and zero spin, embedded in an AGN disk around a MBH of $4\times 10^6 M_\odot$ (see Fig.~\ref{fig:disks}).
For all the three fiducial AGN disks, the migration timescale $t_{\rm mig}$ peaks at $\tilde r \sim 10^2 M_\bullet$,
where $\dot J_{\rm gw}$ becomes comparable with $\dot J_{\rm mig,I}+\dot J_{\rm wind}$. For the $\alpha$ disk case,
$\dot m_{\rm in,0}$ also peaks at $\tilde r\sim 10^2 M_\bullet$ where the gas density $\tilde \rho$
peaks, while $\dot m_{\rm in,0}$ in the  $\beta$ disk is higher than that in the $\alpha$ disk due to higher gas density $\tilde \rho$
in the  $\beta$ disk. Among of the three, the gas density in the TQM disk is the lowest, inside which  $\dot m_{\rm in,0}$ is also the lowest.

\subsection{Mass and spin}
In Section~\ref{subsec:interactions}, we have outlined the calculation of the accretion rate of a sBH
embedded in an AGN disk.  Now we consider the resulting  mass and spin evolution:
\be\label{eq:evol}
\begin{aligned}
  dm_\bullet &= E_0(a) dm_{\rm gas} \ ,\\
  d\mathcal J_\bullet &= L_0(a)m_\bullet dm_{\rm gas} = \frac{L_0(a)}{E_0(a)} m_\bullet dm_\bullet\ ,
\end{aligned}
\ee
where $dm_{\rm gas}$ is the mass element of accreted gas, a fraction $1-E_0(a)$ of which
is converted to radiation escaping to infinity and the remaining fraction $E_0(a)$ is absorbed by the BH,
and $\mathcal J_\bullet=am_\bullet^2$ is the sBH angular momentum.
As a result, we obtain the following simple evolution equations for the mass and the spin
\be\label{eq:1d}
\begin{aligned}
  \dot m_\bullet &= E_0(a) \dot m_{\rm in,0} \ ,\\
  \dot a &= \left[\frac{L_0(a)}{E_0(a)}-2a\right] \frac{\dot m_\bullet}{m_\bullet}\ .
\end{aligned}
\ee
Together with Eq.~(\ref{eq:tmig}), one can solve the migration and the mass and spin evolution of the sBH self-consistently
in the AGN disk. In Fig.~\ref{fig:mass_spin}, we show the mass and spin evolution of sBHs for a duration $10^6$ yr with initial mass $m_{\bullet, \rm ini}=5/10/20 M_\odot$, initial spin $a_{\rm ini}=0$ and
initial location $\tilde r_{\rm ini} = 10^5 M_\bullet$.
In the $\alpha$ disk and the  $\beta$ disk model, the sBHs have successfully migrated into the MBH in less than $10^6$ yr
and  most of mass accumulation and spin amplification happen in the range of $\tilde r \lesssim 10^3 M_\bullet$, where both the migration timescale $t_{\rm mig}$ is relatively  long
and the sBH accretion rate $\dot m_{\rm in,0}$ is relatively high (see Fig.~\ref{fig:mdot_tmig}).
In the TQM disk model, the mass accumulation and the spin increase of the sBH are much slower because of lower sBH accretion rate (see Fig.~\ref{fig:mdot_tmig}). Note that the TQM disk model is not expected to be a good description to
inner parts ($\tilde r\lesssim 10^5 M_\bullet$ for the fiducial disks in Fig.~\ref{fig:disks}) of AGN disks in nature,
as explained in the beginning of Section~\ref{sec:BHs}, so we will focus on $\alpha$ and $\beta$ disks in the following discussion.

\subsection{Distribution of sBHs in the AGN disk}
In this subsection, we focus on the density distribution of sBHs in inner parts of the AGN disk and all  discussions here
largely depend on our previous work \citep{Pan2021}. We will find out how many sBHs are expected in the inner part of the disks where the mass and spin evolution of sBHs become more significant.

As sBHs and stars in the cluster orbiting around the accreting MBH, some of them are captured by the AGN disk and form a disk component. The statistical properties of cluster-component stars and sBHs are encoded in their distribution functions
$f_i(t, \vec x, \vec v)$ ($i=$ star or bh). Following Refs.~\citep{Cohn1978,Cohn1979}, the distribution functions are approximately
functions of the action variables, $f_i\approx f_i(t,E,R)$, with $E:=\phi(\tilde r) -v^2/2$ being the specific binding energy (i.e.,
the positive potential energy minus the kinectic energy) and $R:=J^2/J_c^2(E)$ being the normalized angular momentum, where $J$ is the specific orbital angular momentum and $J_c(E)$ is the specific orbital
angular momentum of a circular orbiter with specific energy $E$.
The distribution functions of cluster-component stars and sBHs are governed
by the orbit-averaged Fokker-Planck equation \citep{Pan2021}
\be\label{eq:FP_disk}
  \mathcal C\frac{\partial f}{\partial t}
  = - \frac{\partial}{\partial E} F_E
  - \frac{\partial}{\partial R} F_R + S\ ,
\ee
where $\mathcal C=\mathcal C(E,R)$ is a normalization coefficient,
$F_{E,R}$ is the flux in the $E/R$ direction
\be\label{eq:flux}
\begin{aligned}
  -F_E &= \mathcal D_{EE}\frac{\partial f}{\partial E} + \mathcal D_{ER}\frac{\partial f}{\partial R} + \mathcal D_E f\ ,\\
  -F_R &= \mathcal D_{RR}\frac{\partial f}{\partial R} + \mathcal D_{ER}\frac{\partial f}{\partial E} + \mathcal D_R f\ ,
\end{aligned}
\ee
with the diffusion coefficients $\{\mathcal D_{EE}, \mathcal D_{RR}, \mathcal D_{ER}\}$,
and the advection coefficients $\{\mathcal D_{E}, \mathcal D_{R}\}$ incorporating two-body scatterings in the cluster and
interactions of stars and sBHs with the AGN disk.
The (negative) source term $S$ arises from stars or sBHs captured by the disk, with
\be
S_{\rm bh} = -\mu_{\rm cap}\mathcal C  \frac{f_{\rm bh}}{t_{\rm mig}^{\rm star}}\frac{m_{\rm star}}{m_{\rm bh}}\ ,\quad
S_{\rm star} = -\mu_{\rm cap}\mathcal C  \frac{f_{\rm star}}{t_{\rm mig}^{\rm star}} \ ,
\ee
where $\mu_{\rm cap}$ is a parameter quantifying the disk capture efficiency and we take $\mu_{\rm cap}=0.1$
as a fiducial value.

For sBHs in the AGN disk, the continuity equation is
\be\label{eq:Nb_r}
\pder[\Sigma_\bullet]{t} + \frac{1}{\tilde r} \pder{t}(\tilde r\Sigma_\bullet v_{r,\bullet}) =
\frac{1}{2\pi\tilde r}\int -S_{\rm bh} \frac{dE}{d\tilde r}\Bigg|_{E=\phi(\tilde r)/2} dR\ ,
\ee
where $\Sigma_\bullet$ is the surface number density of sBHs in the disk,
$v_{r,\bullet}=\tilde r/t_{\rm mig}$ [Eq.~(\ref{eq:tmig})] is the migration velocity in the radial direction,
and the source term on the right-hand side comes
from sBHs captured by the disk.

As an example, we consider a MBH with $M_\bullet=4\times 10^6 M_\odot$ hosting a stellar
cluster consisting of $1 M_\odot$ stars and $10 M_\odot$ sBHs. We initialize the density profiles of stars and sBHs
$n_{\rm star}(r)$ and $n_{\bullet}(r)$
according to the Tremaine's cluster model \citep{Tremaine1994,Dehnen1993} with the total star mass within the influence sphere
of the MBH $\approx M_\bullet$ and the sBH number density is $n_\bullet(r)=10^{-3}n_{\rm star}(r)$.
According to Soltan's argument \citep{Soltan1982}, the average total duration of AGN active phase is $\sim (10^7, 10^9)$ yr,
MBHs should be quiet most of time. To mimick the AGN duty cycle, we evolve the system of the MBH+stellar cluster for $T_0=5$ Gyr without including of interactions of stars and sBHs with the disk, then turn on the disk influence and continue the evolution for a short time $T_{\rm disk}$ \citep[see][for all the details of modelling and calculation]{Pan2021}.

\begin{figure}
  \includegraphics[scale=0.8]{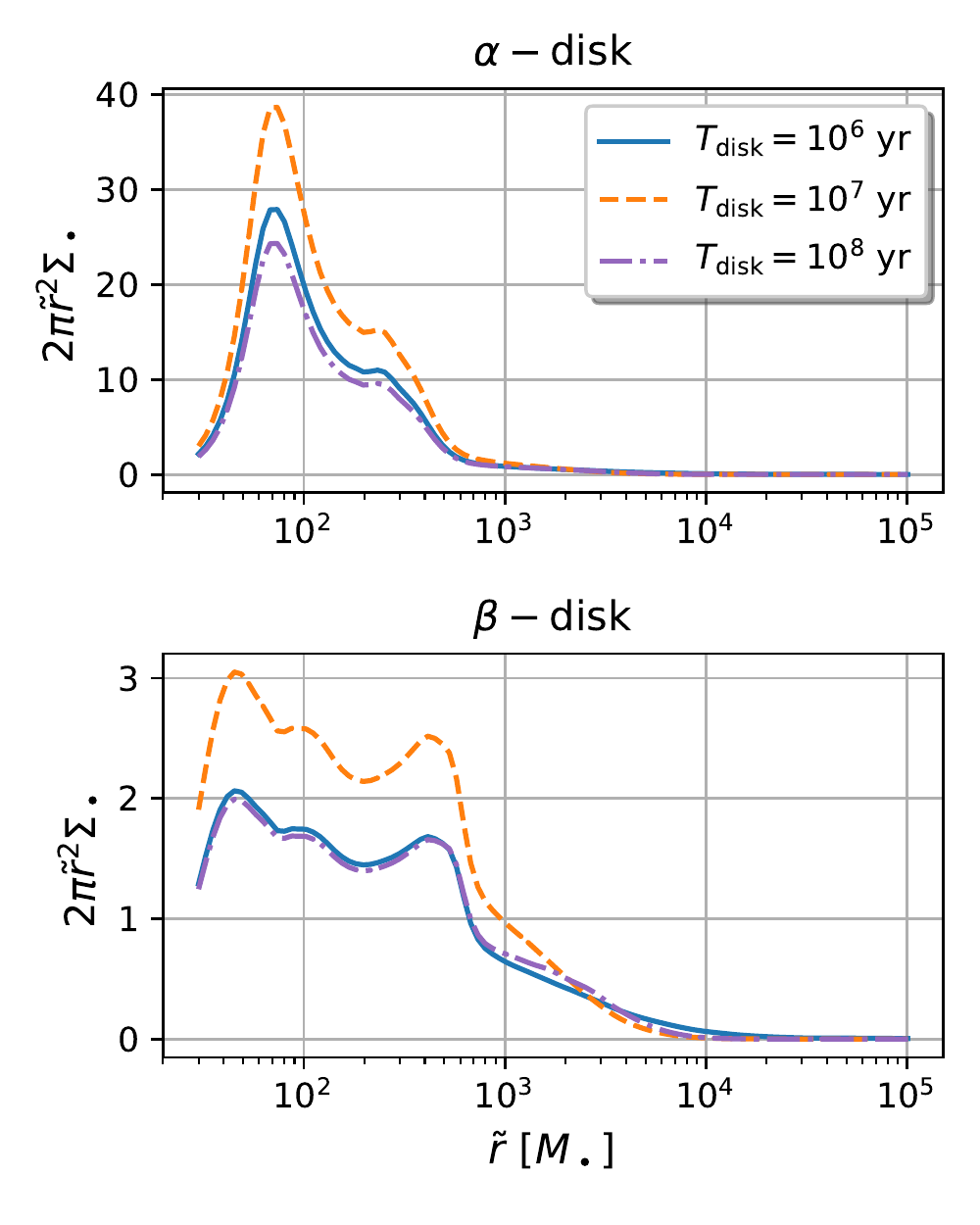}
  \caption{The surface number density $\Sigma_{\bullet}(\tilde r < 10^5 M_\bullet)$ of sBHs on the equator at the end of an AGN phase with duration $T_{\rm disk}$.}
  \label{fig:Nb_r}
\end{figure}

The surface number density $\Sigma_\bullet(r)$ of sBHs captured in the disk is governed by the continuity equation (\ref{eq:Nb_r}).
In Fig.~\ref{fig:Nb_r}, we show the surface number density
$\Sigma_{\bullet}$ of sBHs on the equator at the end of an AGN phase with duration $T_{\rm disk}=10^6/10^7/10^8$ yr.
The total number of sBHs within radius $\tilde r^\star$ is related to the surface number density by
$N_\bullet(<\tilde r^\star) = \int_{< \ln \tilde r^\star} 2\pi r^2\Sigma_\bullet \ d \ln \tilde r $.
In the $\alpha$ disk, we find $\approx (30 - 50)$ sBHs aggregate in the range $\tilde r < 10^3 M_\bullet$
at the end of an AGN phase as a result of a traffic jam.\footnote{\citet{Gilbaum2021} also found the aggregation of
a very similar number of sBHs around the MBH in their feedback-dominated accretion flow of an accreting MBH of mass $10^7 M_\odot$ and accretion rate $0.1 \Medd$, though we have assumed different origins of sBHs embedded in the AGN disk: captured from the stellar cluster v.s. born in the AGN disk.}
In the $\beta$ disk, only $\approx (5 - 8)$ sBHs aggregate in this range
where the migration timescale is much shorter than in the $\alpha$ disk.

Assuming the galactic MBH in Sgr A* went through an AGN phase that ended $10$ Myrs ago \citep{Levin:2003kp},
a number of sBHs should be brought to the vicinity of the MBH as shown above.
If the masses of sBHs are close to $10 M_\odot$, initially the closer ones ($\tilde r \lesssim 200 M_\bullet$) should have been swallowed
by the MBH as a result of GW emissions, while the remaining $\approx (2-12)$ sBHs are expected to be orbiting around the MBH today
with radius $\tilde r < 10^3 M_\bullet$. These remaining sBHs may be promising monochromatic sources detectable by LISA.

\section{Evolution of neutron stars and white dwarfs in an AGN disk}\label{sec:NSs}

\subsection{Neutron stars}
The supercritical accretion onto NSs is similar to that of BHs except with different inner boundary condition and possible effects from magnetic fields. For a BH, the inner boundary condition is naturally an no-torque condition on the horizon.
For a NS, its magnetic field plays an important role, which  has been investigated for decades in the context of
NS accretion in X-ray binaries. The accretion disk  of a NS is expected to be truncated where the magnetic pressure becomes dominant \citep{King2016}. The gas is accelerated or decelerated from approximately Keplerian angular velocity
to the corotation angular velocity by the magnetic stress in a boundary layer between the magnetosphere and the undisrupted accretion disk, then flows along the field lines to the star surface \citep{Ghosh1979a,Ghosh1979b,Ghosh1979c}.
On the other hand, the NS is either spun up or spun down as a result of the angular momentum gain from accreted gas and
the angular momentum loss as magnetically interacting with the accretion disk.

As the  surrounding gas accretes onto the NS,
the  magnetic fields near the star surface  generally decays due to Ohmic dissipation in the heated crust, or being buried under a mountain of the accreted material or leading to the destruction of superconducting vortices in the core of the NS \citep[see  e.g.,][]{Bisnovatyi1974,Konar1997, Istomin2016,Konar2017}. These mechanisms are proposed to explain the fact
that recyled pulsars have substantially weaker surface magnetic fields ($\sim 10^{8-9}$ G) compared to young radio pulsars
($\sim 10^{11-13}$ G) \citep{Manchester2005}.
In particular, in the Ohmic dissipation model, the NS magnetic field decays as the crust is heated up
and the decay stops when the crust sinks into the superconducting core \citep{Konar1997}.
In other decay models,  whether there is a mechanism to stop the field decay is not clear.
In the following discussion, we  adopt the Ohmic heating model and parameterize the NS surface field evolution as
\be
B_{\rm ns}(t) = {\rm max}\left\{B_{\rm ns, ini}\frac{1}{1+t/\tau}, B_{\rm ns, fin}\right\}\ ,
\ee
where $\tau$ is the decay timescale, and $B_{\rm ns, fin}$ is final frozen field strength.
For an  NS accreting gas with Eddington accretion rate $\dot m_{\rm ns}^{\rm Edd}$, $\tau\approx 10^3$ yr and $ B_{\rm ns, fin}\approx (10^9-10^{10})$ G \citep{Konar1997}.

In order to calculate the  accretion rate $\dot m_{\rm ns}$ of a NS with mass $m_{\rm ns}$ and surface magnetic field $B_{\rm ns}$ and embedded in an AGN disk, we first numerically solve the supercritical inflow and outflow
model assuming the background spacetime is  Schwarzschild as in Section~\ref{sec:model},
then truncate the disk at radius $r_{\rm ibd}$ where the magnetic pressure equals the ram pressure \citep[see][for more
detailed calculation]{Ghosh1979a,Ghosh1979b,Ghosh1979c},
and in practice we have
\be
r_{\rm ibd} = {\rm max}\left\{r_{\rm ns}, \xi\left(\frac{B_{\rm ns}^4 r_{\rm ns}^{12}}{2m_{\rm ns}\dot m^2_{\rm in}(r_{\rm ibd})}\right)^{1/7}\right\}\ ,
\ee
where $r_{\rm ns}$ is the NS radius, $\xi$ is a factor of $\mathcal O(1)$  which we take as $\xi=0.52$ following \cite{Ghosh1979c}
and we have assumed a dipolar configuration of the NS magnetic field.
Therefore the NS accretion rate is
\be
 \dot m_{\rm ns} =
 \begin{cases}
   0 \ , \quad\hfill (r_{\rm ibd} = r_{\rm ns}\ {\rm and}\ \Omega_{\rm ns}/\Omega_{\rm K}(r_{\rm ns})>1) \\
   {\rm max}\left\{\dot m_{\rm in}(r_{\rm ibd}), \dot m_{\rm ns}^{\rm Edd} \right\}\ ,\quad\hfill ({\rm otherwise}) \\
 \end{cases}
\ee
considering that the accretion onto the star surface stops when the NS reaches the shedding limit with its angular velocity $\Omega_{\rm ns}$
higher than the  Keplerian angular velocity $\Omega_{\rm K}(r_{\rm ns})$ on the NS surface,
and that the supercritical inflow-outflow model may not hold where the inflow rate falls below the Eddington rate (see Fig.~\ref{fig:sd} and Section~\ref{sec:model}).
The magnetic field expands the effective size of the NS (if $r_{\rm ibd} > r_{\rm ns}$) and therefore accelerates its accretion.

The mass and spin evolution of the magnetized NS depends on the ratio of the NS angular velocity over the disk Keplerian angular velocity at the inner boundary, $\omega_{\rm ns}:=\Omega_{\rm ns}/\Omega_{\rm K}(r_{\rm ibd})$. In the extreme propeller regime ($\omega_{\rm ns}\gg 1$), accretion onto the NS is hindered by the centrifugal barrier and the NS is spun down as interacting with the accretion disk via the magnetosphere \citep{Ghosh1979a,Ghosh1979b,Ghosh1979c}. In the accretion regime ($\omega_{\rm ns} < 1$), the NS is spun up by gas accretion. The equilibrium
is reached at $\omega_{\rm ns}\approx 1$. Following \citet{Ghosh1979c} and \citet{Dai2006}, we
assume the accretion onto the NS is not hindered adruptly in
 the mild propeller regime, and the total torque on the NS is
\be
\frac{d J_{\rm ns}}{dt}= T_{\rm acc} + T_{\rm mag}\ ,
\ee
where $J_{\rm ns}=\frac{2}{5}m_{\rm ns}r_{\rm ns}^2\Omega_{\rm ns}$.
There are various prescriptions of the accretion torque carried by gas falling onto the NS $T_{\rm acc}$
and the magnetic torque $T_{\rm mag}$ exerted on the NS by the accretion disk exterior to the boundary layer.
As modelled by \citet{Ghosh1979a,Ghosh1979b,Ghosh1979c},  the accretion torque is
\be
T_{\rm acc}^{(1)} = T_0:= \dot m_{\rm ns}\sqrt{m_{\rm ns} r_{\rm ibd}}\ ,
\ee
assuming all angular momentum of gas at the inner boundary is absorbed by the NS,
and the total torque is fitted as
\be \label{eq:torque1}
\begin{aligned}
  T_{\rm acc}^{(1)}+T_{\rm mag}^{(1)}
  &= T_0\frac{1.39}{1-\omega_{\rm ns}} \\
  &\times \left\{ 1-\omega_{\rm ns} \left[4.03(1-\omega_{\rm ns})^{0.173}-0.878\right]\right\}\ .
\end{aligned}
\ee
In a revised model \citep{Dai2006}, they are formulated as
\be\label{eq:torque2}
\begin{aligned}
T_{\rm mag}^{(2)} &= \frac{B_{\rm ns}^2r_{\rm ns}^6}{r_{\rm ibd}^3}
\begin{cases}
  \frac{\xi^{7/2}}{3}\left(1-2\omega_{\rm ns} + \frac{2}{3}\omega_{\rm ns}^2 \right)\ , \hfill (\omega_{\rm ns} \leq 1) \\
  \frac{\xi^{7/2}}{3}\left(\frac{2}{3\omega_{\rm ns}}-1\right)\ , \hfill (\omega_{\rm ns} > 1)\
\end{cases} \\
T_{\rm acc}^{(2)} &=
\begin{cases}
  T_0(1-\omega_{\rm ns})\ ,\quad \hfill (r_{\rm ibd} > r_{\rm ns})\\
  T_0 \ , \quad \hfill (r_{\rm ibd} = r_{\rm ns})
\end{cases}
\end{aligned}
\ee
assuming only part of the gas angular momentum is transfered to the NS
via the interaction between magnetic fields and the gas in the boundary layer.
The equilibrium state is reached at $\omega_{\rm ns} = 0.35$ in the first prescription, and at $\omega_{\rm ns} = 0.88$ in the second.

\begin{figure*}
  \includegraphics[scale=0.6]{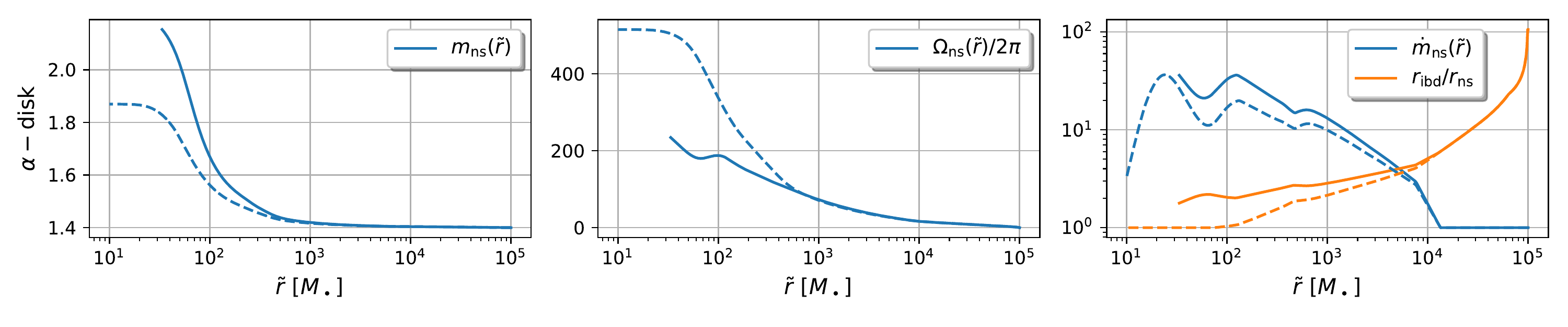}
  \includegraphics[scale=0.6]{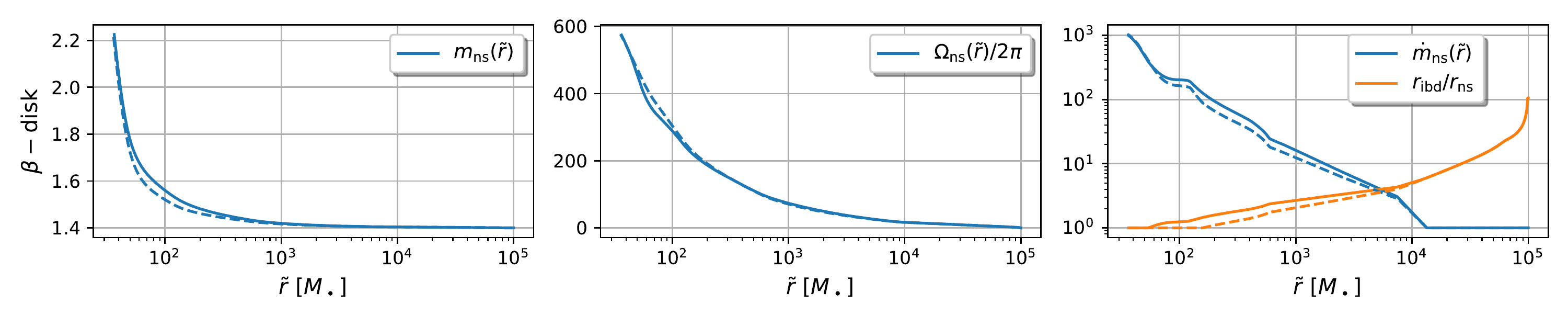}
  \caption{Mass and spin evolution of a NS embedded in an AGN disk assuming the first prescription (Eq.~\ref{eq:torque1}) of the angular momentum exchange.
  Upper row: the mass (left panel), spin frequency $\Omega_{\rm ns}/2\pi\ [\rm Hz]$ (middle panel), accretion rate $\dot m_{\rm ns}\ [\dot m_{\rm ns}^{\rm Edd}]$ and the inner radius the accretion disk
  $r_{\rm ibd}\ [r_{\rm ns}]$ (right panel) of a NS in the fiducial $\alpha$ disk with initial mass $m_{\rm ini}=1.4 M_\odot$ and initial position $\tilde r_{\rm ini}=10^5 M_\bullet$, initial magnetic field $B_{\rm ns, ini}=10^{12}$ G,
  and finally frozen magnetic field $B_{\rm ns, fin}=10^{10}$ G (solid lines) or $10^9$ G (dashed lines).
  Lower row: similar to the upper row except with the fiducial $\beta$ disk.}
  \label{fig:ns_evol1}
\end{figure*}

\begin{figure*}
  \includegraphics[scale=0.6]{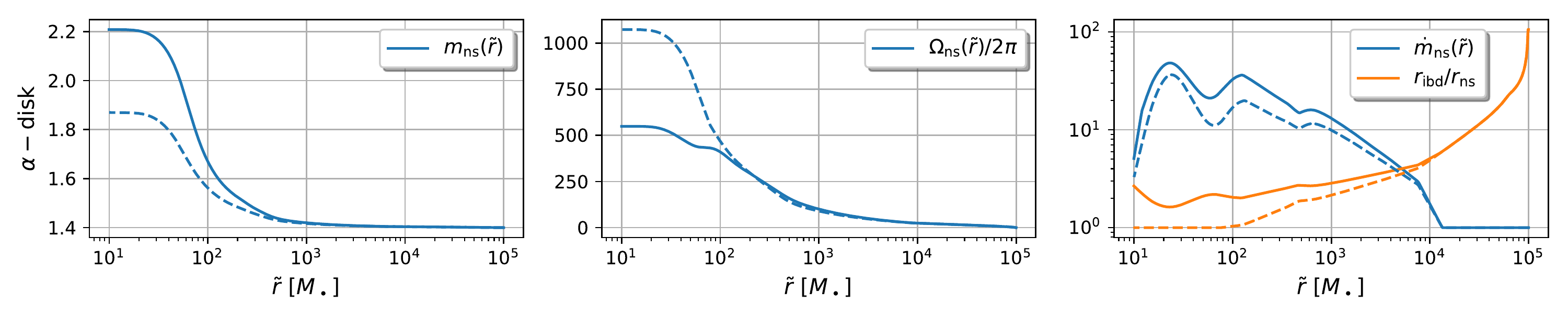}
  \includegraphics[scale=0.6]{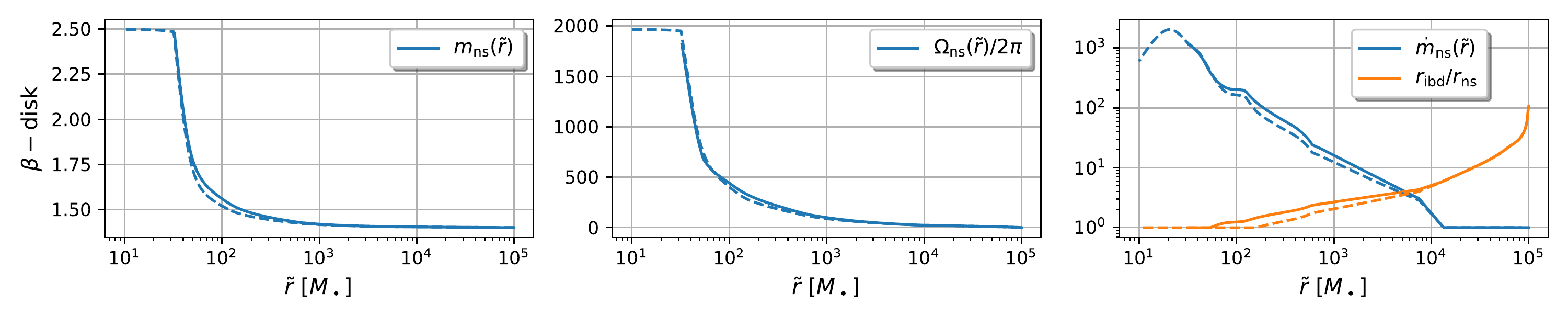}
  \caption{Similar to Fig.~\ref{fig:ns_evol1} except with the second prescription (Eq.~\ref{eq:torque2}) for the angular momentum exchange.}
  \label{fig:ns_evol2}
\end{figure*}

We consider the evolution of NSs in an AGN with initial mass $m_{\rm ns,ini}= 1.4 M_\odot$, radius $r_{\rm ns} = 13$ km,
initial location $\tilde r_{\rm ini} = 10^5 M_\bullet$ and initial surface magnetic field $B_{\rm ns, ini}=10^{12}$ G.
We also take $\tau=10^3$ yr and $B_{\rm ns, fin}=10^{10}$ G or $10^{9}$ G to mimick
the field decay in the Ohmic dissipation model \citep{Konar1997}.
We show the evolution in Figs.~\ref{fig:ns_evol1} and \ref{fig:ns_evol2} with the two prescriptions of angular momentum
exchange, respectively. In both the $\alpha$ disk and the $\beta$ disk,
it takes the NS slightly more than $10^6$ yr to migrate from its initial location to the MBH,
where  the NS surface magnetic field decays from $10^{12}$ G to $10^{10}$ G
in $\sim 10^5$ yr as migrating from $\tilde r =10^5 M_\bullet$ to $10^4 M_\bullet$.
Massive NSs are stable only if they are below
the collapse limit which is $\approx (2.1-2.5) M_\odot$ depending on the NS rotation angular velocity
\citep{Rezzolla2018}, and we simply assume a linear dependence
\be\label{eq:Mmax}
M_{\rm ns, max}/M_\odot = 2.1 + 0.4 \Omega_{\rm ns}/\Omega_{\rm K}(r_{\rm ns})\ .
\ee

With the first prescription of angular momentum exchange (Eq.~\ref{eq:torque1}) and the fiducial $\alpha$ AGN disk,
the strong magnetic field $B_{\rm fin} = 10^{10}$ G always truncates the NS accretion flow and keeps the NS in a slow-rotation
state with angular velocity $\Omega_{\rm ns}\approx 0.35 \Omega_{\rm K}(r_{\rm ibd})$, and the NS collapses with mass $\approx 2.2 M_\odot$ and spinning frequency $\approx 200$ Hz.
For the weak frozen field strength case ($B_{\rm fin} = 10^9$ G), the NS accrete rate is lower and the NS does not reach the collapse limit with final mass $\approx 1.8 M_\odot$. In the fiducial $\beta$ disk in which the gas density is higher than in the $\alpha$ disk, thus the NS accretion rate is also higher. As a result, the NS collapses with final mass $\approx 2.2 M_\odot$ for either $B_{\rm fin}=10^{10}$ G or $10^9$ G.

With the second prescription of angular momentum exchange (Eq.~\ref{eq:torque2})  and the fiducial $\alpha$ AGN disk,
the strong magnetic field $B_{\rm fin} = 10^{10}$ G again always truncates the NS accretion flow and
keeps the NS rotating with a higher rate ($\Omega_{\rm ns}\approx  0.88 \Omega_{\rm K}(r_{\rm ibd})$)
than in the first prescription and the NS finally grows to $\approx 2.2 M_\odot$ but does not collapse
because it is still below the spin-enhanced collapse limit.
For the weak frozen field strength case ($B_{\rm fin} = 10^9$ G), the NS accretion rate is lower and
the NS finally grows to $\approx 1.8 M_\odot$ without collapse.
In the fiducial $\beta$ disk, the NS accretion rate is higher as explained in the previous paragraph.
With the strong frozen magnetic field $B_{\rm fin}=10^{10}$ G, the NS grows to and
collapses with mass $\approx 2.45 M_\odot$ and nearly
maximal spin ($\Omega_{\rm ns}\approx 0.9\Omega_{\rm K}(r_{\rm ns})$).
With the weak frozen magnetic field $B_{\rm fin}=10^{9}$ G, the NS accretion rate is lower while is spun up faster,
thus the NS first reaches the shedding limit ($\Omega_{\rm ns}=\Omega_{\rm K}(r_{\rm ns})$) when the NS accretion is largely
suppressed. As a result, the NS does not collapse with the final mass slightly below the maximal mass, which is $2.5 M_\odot$
in our simple prescription.

\begin{figure}
  \includegraphics[scale=0.85]{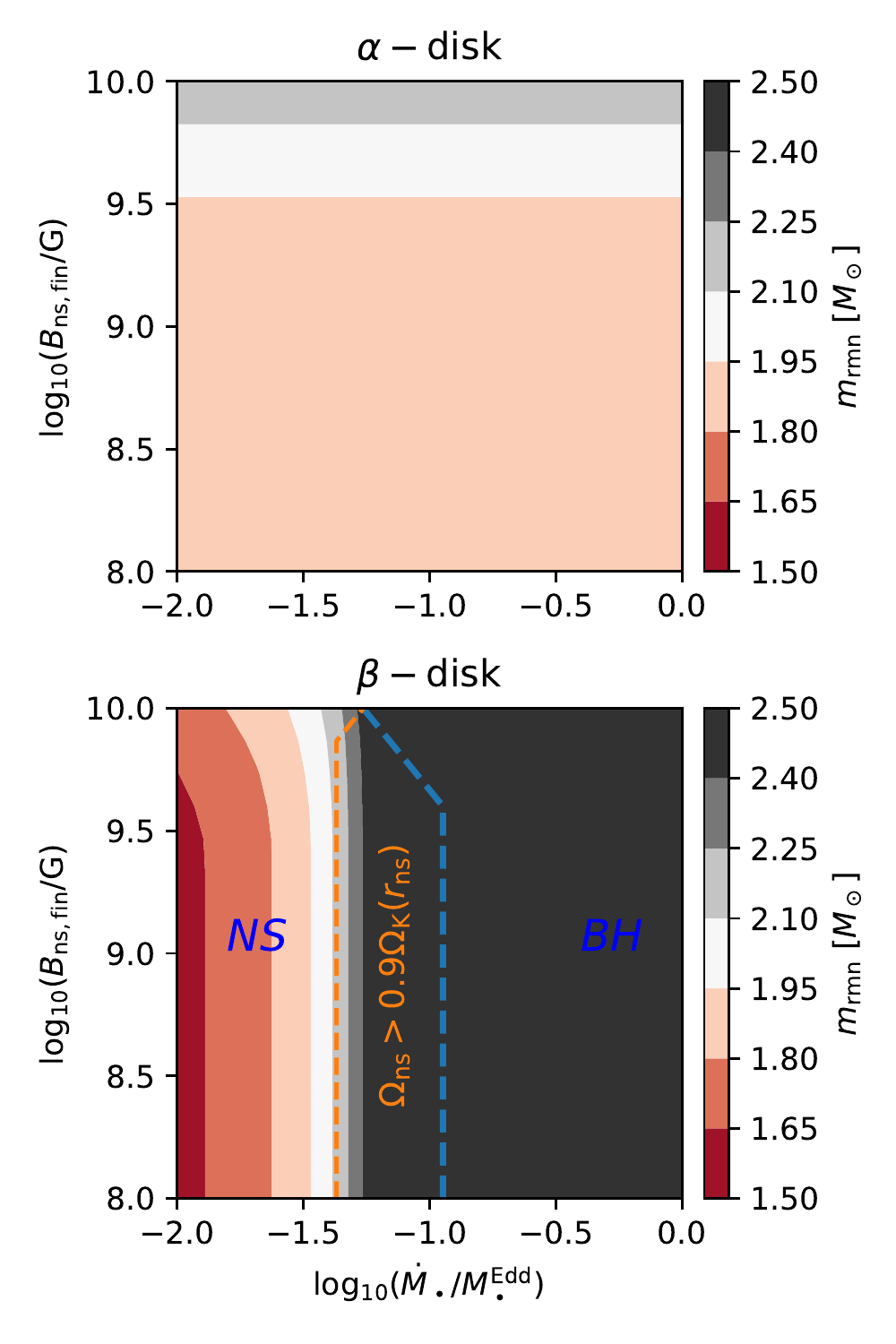}
  \caption{Final fates of NSs with initial mass $1.4 M_\odot$, initial location $\tilde r=10^5 M_\bullet$,
  in an AGN disk of a MBH of mass $M_\bullet=4\times10^6 M_\odot$ and accretion rate $\dot M_\bullet$.
  For $\alpha$ disk model, the NS accretion rate is low, the NS does not grow to the collapse limit before merging with the central MBH.
  For $\beta$ disk model with accretion rate $\dot M_\bullet \gtrsim 0.1 \Medd$, the NS accumulates sufficient mass and collapses with
  final mass $m_{\rm rmn}\approx 2.5 M_\odot$ (right side of the blue dashed line).
  For $\beta$ disk model with accretion rate $\dot M_\bullet \lesssim 0.1 \Medd$, the NS does not collapse because its final mass is slightly below the spin-enhanced collapse limit (left side of the blue dashed line). Between the two dashed lines is the parameter space where the NS finally becomes a massive
  and  fast rotating NS.}
  \label{fig:ns_fate}
\end{figure}

We conclude this section by summarizing the final fate of a NS with initial mass $1.4 M_\odot$ and zero initial spin,
captured by an AGN disk around a MBH of mass $M_\bullet=4\times10^6$ at radius $\tilde r=10^5 M_\bullet$.
As an illustrating example, we assume the prescription of angular momentum exchange (\ref{eq:torque2}).
In Fig.~\ref{fig:ns_fate}, we show the influence of  different AGN disk models, MBH accretion rates
and the frozen NS magnetic field strength $N_{\rm ns, fin}$.
For $\alpha$ disk model in which the gas density is relatively low,
the accretion rate is low so that the NS does not grow to the collapse limit before merging with the central MBH.
For $\beta$ disk model with accretion rate $\dot M_\bullet \gtrsim 0.1 \Medd$, the NS accumulates sufficient mass and collapses with final mass $m_{\rm rmn}\approx 2.5 M_\odot$.
For lower accretion rate $\dot M_\bullet \lesssim 0.1 \Medd$, the NS does not collapse even if its final mass is marginally below the spin-enhanced collapse limit.
To summarize, magnetic fields play an important role in the mass accumulation and spinning up of accreting NSs embedded in AGN disks: stronger magnetic fields lead to higher NS accretion rate and lower NS spinning rate,
resulting in heavier NSs and easier NS collapses. The AGN disk structure, especially its gas density, also makes a big difference in the fate of NSs.

\subsection{WDs}
The evolution of WDs accreting gas from the stellar wind of a companion star
has been extensively studied previously, which can give rise to very rich phenomena depending on the accretion rate
\citep[see e.g.,][]{Gallagher1978,Fujimoto1982,Iben1982,Hachisu1996,Cassisi1998,Wolf2013,Chomiuk2020}.
A novae is ignited due to unstable burning of hydrogen on the WD surface
if $\dot m_{\rm wd} < \dot m_{\rm wd, stable}$, where $\dot m_{\rm wd, stable}\approx  (2\times10^{-8}-4\times10^{-7}) M_\odot$/yr
depending on the WD mass is the minimal accretion rate for sustaining stable burning. For accretion rates slightly above $\dot m_{\rm wd, stable}$, the stable hydrogen burning rate is equal to its accretion rate, so the WD continually accumulates mass until
reaching the Chandrasekhar limit or the shedding limit. For even higher accretion rates $\gtrsim 3\dot m_{\rm wd, stable}$, the hydrogen burning rate is lower than its accretion rate, thus the accreted hydrogen piles up on the WD surface as a radially expanding envelope (similar to that in red giants) until optically thick winds
evetually slow down the net accretion.

\begin{figure}
  \includegraphics[scale=0.7]{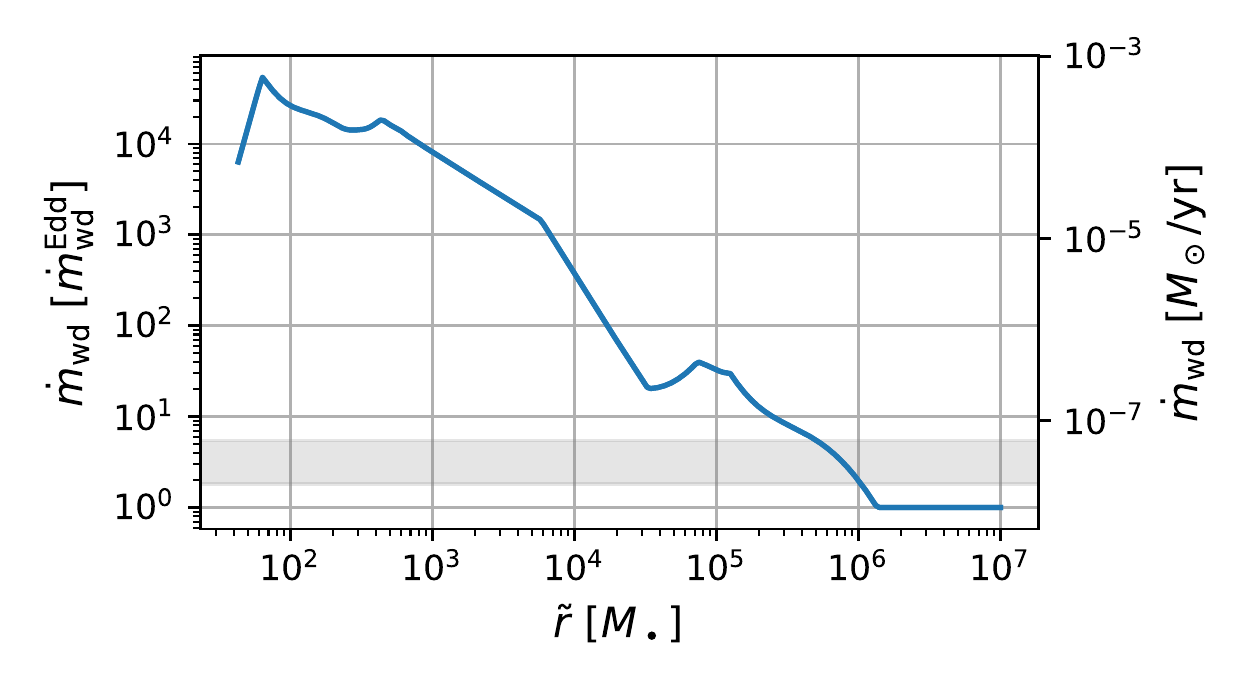}
  \caption{Accretion rates of WDs with mass $m_{\rm wd}=0.5 M_\odot$ and radius $r_{\rm wd}=10^4$ km,
  located at different radii in the fiducial $\alpha$ AGN disk. The gray band is the stable burning regime where the hydrogen buring rate matches the accretion rate.}
  \label{fig:mdot_wd}
\end{figure}

We now examine the accretion rates of WDs placed in various locations of an AGN disk.
Surface magnetic fields of WDs are generally weak ($ < 10^5$ G) \citep{Landstreet2012},
therefore have little impact on the accretion flows of WDs. For a WD embedded in an AGN disk,
the star accretion rate $\dot m_{\rm wd}$ is generally much higher compared with sBHs or
NSs of similar mass because of its much larger size $r_{\rm wd}$, with
\be
\dot m_{\rm wd} ={\rm max}\left\{\mdoto\times  r_{\rm wd}/r_{\rm obd}, \dot m_{\rm wd}^{\rm Edd}\right\}\ .
\ee
In Fig.~\ref{fig:mdot_wd},
we show the accretion rates of WDs with mass $m_{\rm wd}=0.5 M_\odot$ and radius $r_{\rm wd}=10^4$ km, assuming the fiducial $\alpha$ AGN disk. For WDs located in a narrow annulus $\tilde r \approx 2\times 10^6 M_\bullet$, their accretion rates
fall in the stable hydrogen burning regime $(1-3)\dot m_{\rm wd, stable}\approx(2-6)\times 10^{-8} M_\odot/$yr (for $m_{\rm wd}=0.5 M_\odot$) \citep{Wolf2013}. For WDs located exterior to this annulus,  their accretion rates is lower
than $\dot m_{\rm wd, stable}$, thus these WDs do not grow
because accumulated mass is blown away during novae eruptions. For WDs located at $\tilde r < 10^6 M_\bullet$,
the accretion rates are above $3\dot m_{\rm wd, stable}$,
and the accreted gas piles up on the WD surface causing an expanding red-giant-like structure
with $r_{\rm rg}\gg r_{\rm wd}$.  In a short period of time $\delta t \ll m_{\rm wd}/\dot m_{\rm wd}$,
the ``red giant" is spun up to the shedding limit $\sqrt{m_{\rm wd}/r_{\rm rg}^3}$ and the accretion stops.
Therefore it seems difficult for WDs embedded in AGN disks to grow to the Chandrasekhar limit via accretion.
This picture is also true for other two AGN disk models.

\section{Summary}\label{sec:summary}
AGN accretion disks have been proposed as  promising sites for producing
both mergers of stellar-mass compact objects and extreme mass ratio inspirals.
For compact objects captured into an AGN disk,  gas accretion generically happens and affects the long-term evolution of these objects,
which has not been throughly understood and discussed in previous studies.
In this paper, we construct a relativistic supercritical inflow-outflow model of BHs, based on previous Newtonian models and
GRMHD simulations. In this model, the gas inflow rate $\dot m_{\rm in} (r)$ is an increasing function of radius $r$
because of the existence of outflows. Apart from having  radius-dependent inflow rates, inflows in this model are very similar to canonical ADAFs \citep{Narayan1994,Narayan1997},
both of which are radiation inefficient and advection dominated.
As an application, we considered the speculated EM counterpart of GW190521, an optical flare of luminosity $\sim 10^{45}$ erg/s,
which has been interpreted as the radiation from supercritical accretion of the remnant BH of mass $\sim 10^2 M_\odot$
in an AGN disk with an Bondi accretion rate $\approx 2\times 10^5 \dot M_{\rm BBH}^{\rm Edd}$ and an radiation efficiency
$\eta_{\rm rad}\approx 0.1$ \citep{Graham2020}.  There is an upper limit of the disk luminosity $\lesssim 10 L_{\rm Edd}$
in the supercritical inflow-outflow model, which admits no solution to such  high luminosity.
To avoid possible model bias, we also invoke the model of supercritical accretion without outflow \citep{Beloborodov1998},
which admits a much lower radiation efficiency $\eta_{\rm rad}\approx 3.5\times 10^{-5}$ (Eq.~\ref{eq:eta_in}).
Together with the energy argument, we conclude that the supercritical accretion onto the remnant BH seems unlikely to be the origin of  such a bright optical flare.

We have applied the supercritical inflow-outflow model to study the evolution of sBHs embedded in AGN disks.
For such a sBH, surrounding gas within its gravitational influence spheres tends to flow toward it,
then circularizes and forms a disk-like profile due to non-zero angular momentum carried by the gas w.r.t the sBH.
The outer boundary $r_{\rm obd}$ and the gas inflow rate at the outer boundary $\dot m_{\rm in}(r_{\rm obd})$
are specified by the circularization radius and the Bondi accretion rate with environmental corrections, respectively.
We find that the inflow rate at the outer boundary $\dot m_{\rm in}(r_{\rm obd})$ is in general highly super-Eddington,
while the sBH accretion rate $\dot m_{\bullet, 0}$ is only a small portion of $\dot m_{\rm in}(r_{\rm obd})$, which turns out to be mildly
super-Eddington in most cases.
As a result, the majority of sBHs that are captured onto the AGN disk and migrate into the MBH within the AGN
disk lifetime only grow by a small fraction due to accretion.

We also applied the supercritical inflow-outflow model on studying NSs and WDs in AGN disks,
taking into account corrections from star sizes and star magnetic fields.
Accretion rates of embedded WDs rates are usually higher than that of BHs of similar mass because of much larger star sizes. However, WDs are spun up more efficiently to the shedding limit when the accretion onto star surface ceases because of angular momentum barrier,
therefore it is hard for WDs to grow to the Chandrasekhar limit via accretion.
For NSs, the surface magnetic fields may play an important role in both
spinning up and accelerating the accretion. If NS magnetic fields are sufficiently strong to guide the accretion flow  outside the NS surface and consequently keep the NS in a slow rotation state while accreting gas,  it is possile for NSs to grow to the collapse limit via accretion. This opens up possibilities to form mass-gap-EMRIs that are detectable by LISA, whose event rate requires a separate study.

\section*{Acknowledgements}
Z.P. and H.Y. are supported by the
Natural Sciences and Engineering Research Council of
Canada and in part by Perimeter Institute for Theoretical
Physics. Research at Perimeter Institute is supported in part
by the Government of Canada through the Department of Innovation,
Science and Economic Development Canada and by the Province of
Ontario through the Ministry of Colleges and Universities.

\bibliography{ms}

\end{document}